\documentclass[prl,amssymb,amsmath,amsfonts,nofootinbib,reprint,showpacs,longbibliography]{revtex4-1}

\usepackage{graphicx}
\usepackage{lmodern}
\usepackage{paralist}
\usepackage{amsmath,amssymb}
\usepackage{mathrsfs}
\usepackage{amsfonts}
\usepackage[utf8]{inputenc}
\usepackage{url}
\usepackage[breaklinks, plainpages=false, colorlinks=true, anchorcolor=cyan, linkcolor=magenta, citecolor=cyan, urlcolor=violet, bookmarks=false]{hyperref}
\usepackage[dvipsnames]{xcolor}
\usepackage{multirow}
\usepackage[normalem]{ulem}
\usepackage{lipsum}

\usepackage{marvosym}
\usepackage{enumerate}
\usepackage{color,soul}
\usepackage{acronym}

\usepackage{bm}
\usepackage{color}
\usepackage{commath}
\allowdisplaybreaks
\usepackage{multirow}

\def\TEOBResumSDali{\texttt{TEOBResumS-Dalí}}

\def\c3{c_{\rm{N}^3\rm{LO}}}

\def\alphalm0{\alpha_{\ell m 0}}
\def\taulm0{\tau_{\ell m 0}}
\def\omegalm0{\omega_{\ell m 0}}
\def\bosch{\texttt{pTEOBResumS-Dalí}}

\newacro{adm}[ADM]{Arnowitt-Deser-Misner}
\newacro{bbh}[BBH]{binary black hole}
\newacro{bh}[BH]{black hole}
\newacroplural{bh}[BHs]{black holes}
\newacro{bhns}[BHNS]{black hole-neutron star}
\newacro{bhpt}[BHPT]{black hole perturbation theory}
\newacro{bns}[BNS]{binary neutron star}
\newacro{bf}[BF]{Bayes' factor}
\newacro{cbc}[CBC]{compact binary coalescence}
\newacro{ce}[CE]{Cosmic Explorer}
\newacro{da}[DA]{data analysis}
\newacro{et}[ET]{Einstein Telescope}
\newacro{eob}[EOB]{Effective-One-Body}
\newacro{eom}[EOM]{equations of motion}
\newacro{fd}[FD]{frequency domain}
\newacro{fft}[FFT]{Fast Fourier transform}
\newacro{gw}[GW]{gravitational-wave}
\newacroplural{gw}[GWs]{Gravitational-waves}
\newacro{gr}[GR]{general relativity}
\newacro{grb}[GRB]{gamma-ray burst}
\newacro{grhd}[GRHD]{general-relativistic hydrodynamics}
\newacro{gwosc}[GWOSC]{Gravitational Wave Open Science Center}
\newacro{gwtc1}[GWTC-1]{the first gravitational-wave transients catalog}
\newacro{gsf}[GSF]{Gravitational Self Force}
\newacro{hm}[HM]{Higher mode}
\newacroplural{hm}[HMs]{Higher modes}
\newacro{ifo}[IFO]{interferometer}
\newacro{imr}[IMR]{inspiral-merger-ringdown}
\newacro{im}[IMR]{inspiral-to-merger}
\newacro{kagra}[KAGRA]{Kamioka Gravitational Wave Detector}
\newacro{ligo}[LIGO]{Laser Interferometer Gravitational-Wave Observatory}
\newacro{lisa}[LISA]{Laser Interferometer Space Antenna}
\newacro{lr}[LR]{Light Ring}
\newacro{lso}[LSO]{Last Stable Orbit}
\newacro{lvc}[LVC]{LIGO-Virgo Collaboration}
\newacro{lvk}[LVK]{LIGO-Virgo-Kagra Collaboration}
\newacro{lo}[LO]{leading order}
\newacro{ns}[NS]{neutron star}
\newacroplural{ns}[NSs]{neutron stars}
\newacro{nr}[NR]{numerical relativity}
\newacro{nqc}[NQC]{Next-to-quasicircular corrections}
\newacro{nlo}[NLO]{next-to-leading order}
\newacro{nnlo}[NNLO]{next-to-next-to-leading order}
\newacro{n3lo}[N3LO]{next-to-next-to-next-to-leading order}
\newacro{n4lo}[N3LO]{next-to-next-to-next-to-next-to-leading order}
\newacro{ode}[ODE]{Ordinary Differential Equation}
\newacroplural{ode}[ODEs]{Ordinary Differential Equations}
\newacro{pn}[PN]{post-Newtonian}
\newacro{pm}[PM]{post-Minkowskian}
\newacro{pe}[PE]{parameter estimation}
\newacro{psd}[PSD]{power spectral density}
\newacro{pa}[PA]{post-adiabatic}
\newacro{qnm}[QNM]{quasi-normal mode}
\newacro{qc}[QC]{quasi-circular}
\newacro{snr}[SNR]{signal-to-noise ratio}
\newacro{spa}[SPA]{stationary-phase approximation}
\newacro{sxs}[SXS]{Simulating eXtreme Spacetimes}
\newacro{td}[TD]{time domain}
\newacro{ng}[NG]{Next Generation}
\newacro{sm}[SM]{Supplementary Material}

\definecolor{cyan}{rgb}{0,0.9,0.9}
\definecolor{orange}{rgb}{0.9,0.5,0}
\definecolor{magenta}{rgb}{1,0,1}
\definecolor{purple}{rgb}{0.8,0.4,0.8}
\definecolor{gray}{rgb}{0.8242,0.8242,0.8242}
\definecolor{dodgerblue}{rgb}{0.12, 0.56, 1.0}

\newcommand{\event}[0]{GW250114}

\newcommand{\bilby}[0]{\texttt{Bilby}}
\newcommand{\dynesty}[0]{\texttt{dynesty}}

\begin{document}

\title{From Source Properties to Strong-Field Tests: a multipronged analysis of GW250114 with an effective one-body model for generic orbits}
\author{Koustav \surname{Chandra}${}^{1,2}$} \email{kbc5795@psu.edu}
\author{Rossella \surname{Gamba}${}^{2,3}$} \email{rgamba@berkeley.edu}
\author{Danilo \surname{Chiaramello}${}^{4,5}$}
\affiliation{${}^{1}$ Max Planck Institute for Gravitational Physics (Albert Einstein Institute), Am M\"uhlenberg 1, 14476 Potsdam, Germany}
\affiliation{$^{2}$Institute for Gravitation \& the Cosmos, Department of Physics \& Department of Astronomy and Astrophysics, The Pennsylvania State University, University Park, Pennsylvania 16802, USA}
\affiliation{${}^{3}$ Department of Physics, University of California, Berkeley, CA 94720, USA}
\affiliation{${}^{4}$ Dipartimento di Fisica, Universit\`a di Torino, Via P. Giuria 1, 10125 Torino, Italy}
\affiliation{${}^{5}$ INFN Sezione di Torino, Torino, 10125, Italy}

\begin{abstract}
We present a detailed analysis of \event, the loudest gravitational-wave signal observed to date, using a waveform model capable of describing binary black holes in generic
(eccentric and precessing) orbits. Our analysis builds on LIGO-Virgo-KAGRA (LVK)'s results, finding that the source is consistent at a probability of $\geq 96\%$
with the merger of two first-generation, nearly equal-mass, low-spin black holes, forming a remnant within the pair-instability mass gap.  
The signal's high signal-to-noise ratio (\(\gtrsim 75\)) enables the detection of the subdominant \((4,\pm4)\) multipoles, whose presence we confirm with
higher evidence than previously reported by the LVK. Restricting the analysis even to post-peak data yields \(\log_{10}\mathcal{B}\gtrsim 1\) in favor of models including the \((4,\pm4)\) mode,
demonstrating that this contribution remains detectable well into the post-merger phase.
We further perform three independent tests of general relativity, complementary to those performed by the LVK: a modified residual analysis confirms that our
semi-analytical model fully describes the signal without detectable discrepancies; a subdominant mode test finds that the amplitude of the \((4,\pm4)\)
multipoles agrees with general-relativistic expectations; and a parameterised analysis of the plunge-merger-ringdown regime recovers the GR expectation within
the 50\% credible region for the remnant mass and spin, and within the 90\% interval for the \((2,\pm2)\) peak amplitude. Collectively, these results reinforce GW250114
as a landmark event for a precision test of gravity.
\end{abstract}
\date{\today}
\maketitle

\acresetall


\noindent\textbf{Introduction:}
With an exceptional network \ac{snr} of $\gtrsim 75$, 
GW250114\_082203 (\event, hereafter)—the loudest \ac{gw} signal of the first decade of \ac{gw} astronomy—offers unprecedented statistical power for precision studies of \ac{bbh} mergers
and for performing the most stringent \ac{gw}-based tests of \ac{gr} to date~\citep{LIGOScientific:2025epi, LIGOScientific:2025obp, Kumar:2025jio, Andres-Carcasona:2025bni}. 
Such a high-\ac{snr} signal exemplifies the remarkable sensitivity achieved by the ground-based \ac{gw} detectors and the opportunities that future facilities
promise~\citep{Cahillane:2022pqm, Capote:2024rmo}, particularly in the context of multi-phase tests of \ac{gr}~\citep{Krishnendu:2021fga, Yunes:2025xwp}. 
However, the statistical precision enabled by such signals places stringent demands on all aspects of the analysis.  
In particular, waveform accuracy becomes critical: even small modelling inaccuracies can bias the inferred source properties or masquerade as apparent
deviations from \ac{gr}~\citep{Hu:2022bji, Chandramouli:2024vhw, Chandra:2024dhf, Gupta:2024gun, Dhani:2024jja, LIGOScientific:2025rsn, LIGOScientific:2025cmm}.  
Similarly, neglecting physical effects such as orbital eccentricity or spin-induced precession could in principle bias parameter estimates for certain sources. 

The \ac{lvk} discovery and companion analyses established \event\ as a quasi-circular, low-spin \ac{bbh} merger with source-frame component masses 
\(m_{1}=33.6^{+1.2}_{-0.8}\,M_\odot\) and 
\(m_{2}=32.2^{+0.8}_{-1.3}\,M_\odot\), 
and spins \(\chi_{1},\chi_{2} \leq 0.24,0.26\) (90\% credible intervals)~\citep{LIGOScientific:2025epi, LIGOScientific:2025obp}.  
These estimates were shown to be robust across four state-of-the-art non-eccentric \ac{bbh}
waveform models~\citep{Varma:2019csw, Pompili:2023tna, Khalil:2023kep, vandeMeent:2023ols, Ramos-Buades:2023ehm, Pratten:2020ceb, Colleoni:2024knd}. 
Using eccentric aligned-spin models, the \ac{lvk} further constrained the orbital eccentricity to \(e \leq 0.03\) at 13.33\,Hz~\citep{Gamboa:2024hli, Nagar:2024dzj}, ascertaining the signal's non-eccentric origin.  
Their post-inspiral analyses reported positive evidence for the \((4,\pm 4)\) subdominant mode (\(\log_{10}\mathcal{B}=0.54^{+0.18}_{-0.18}\)) and demonstrated that the post-peak emission is fully consistent with the Kerr \(220\) \ac{qnm} and its first overtone~\citep{Carullo:2019flw, Isi:2019aib}.  
Residual analyses using best-fit waveforms and \ac{imr} consistency tests demonstrated agreement with \ac{gr}, with the latter yielding constraints
comparable to those obtained by hierarchically combining events in GWTC-4~\citep{LIGOScientific:2025tgr} and proving that the event
satisfied Hawking's area increase law~\citep{Hawking:1971tu}.
Finally, parameterised plunge-merger-ringdown tests within the \texttt{pSEOBNR} framework found both the \(220\) and \(440\)
\acp{qnm} consistent with \ac{gr}~\citep{Brito:2018rfr, Ghosh:2021mrv, Maggio:2022hre, Pompili:2025cdc}, while inspiral-phase tests based on \ac{pn}
deviation parameters~\citep{Agathos:2013upa} and principal-component analyses~\citep{Saleem:2021nsb}
reinforced that GW250114 is a textbook example of a \ac{gr}-consistent \ac{bbh} signal.


Here we extend the \ac{lvk} analyses by studying \event\ with \TEOBResumSDali~\citep{Gamba:2024cvy, Nagar:2024oyk, Albanesi:2025txj}, a state-of-the-art \ac{eob}
model capable of describing generic \ac{bbh} orbits. We quantify the statistical evidence of the various physical effects modelled (eccentricity, spin precession or 
a combination of the two), demonstrate decisively the presence of the \((4,\pm4)\) multipoles, and perform three independent tests of \ac{gr} complementary to those presented by the \ac{lvk}.\\

\noindent\textbf{\ac{gr} and beyond-\ac{gr} models:}
A general relativistic \ac{bbh} is intrinsically characterised by nine parameters:
the mass ratio $q=m_1/m_2 \geq 1$, the two (dimensionless) spins $\boldsymbol{\chi}_{1,2}$,
eccentricty $e_0$ and mean anomaly $l_0$ at a reference frequency \(f_{\rm ref}\).
While the total mass does not affect the intrinsic dynamics, it determines the \ac{gw} frequency and the
signal's loudness in physical units, making it relevant for detection and source property measurement. 

\TEOBResumSDali\ semi-analytically models the radiation multipoles or \ac{gw} modes \(h_{\ell m}(t)\) of a \ac{bbh}
system through a two-stage approach within the \ac{eob} framework~\citep{Buonanno:1998gg, Buonanno:2000ef}~\footnote{Our model version includes the \((\ell,m)=(2,\pm2), (2,\pm1), (3,\pm3), (4,\pm4) \) multipoles}.
The inspiral-plunge phase of the modes, \(h_{\ell m}^{\mathrm{insp-plunge}}(t)\), is computed from the \ac{eob} dynamics
via factorized and resummed \ac{pn} expansions and \ac{nr} calibrations, which capture the non-quasicircular effects
during the final plunge~\citep{Damour:2008gu, Damour:2007xr, Nagar:2024oyk}.
This is then smoothly connected to a phenomenologically constructed merger-ringdown template~\citep{Damour:2014yha, DelPozzo:2016kmd} 
\(h_{\ell m}^{\mathrm{merger-RD}}(t)\) at mode-dependent matching times \(t_{\ell m}^{\mathrm{match}}\)~\citep{Nagar:2020pcj}:
\begin{equation}
    h_{\ell m}(t) = 
        \begin{cases}
            h_{\ell m}^{\mathrm{insp-plunge}}(t) & t < t_{\ell m}^{\mathrm{match}} \\
            h_{\ell m}^{\mathrm{merger-RD}}(t) & t \geq t_{\ell m}^{\mathrm{match}}~.
    \end{cases}
\end{equation}
where the latter combines inputs from \ac{bh} perturbation theory and \ac{nr} simulations. \TEOBResumSDali~differs from other \ac{eob}-based models (e.g. those presented in Refs.~\citep{Pompili:2023tna, Khalil:2023kep, vandeMeent:2023ols, Ramos-Buades:2023ehm, Gamboa:2024imd, Gamboa:2024hli}) in the choices made in its construction, from the resummations adopted to -- most prominently -- the treatment of non-quasicircular corrections and the merger-ringdown attachment procedure.
For further details, we refer the reader to \citet{Gamboa:2024hli, Nagar:2024oyk}\\

To probe potential beyond-\ac{gr} effects within the parameterised \ac{imr} framework, we employ two complementary approaches:  
\begin{inparaenum}[(1)]
    \item introduce fractional mode–amplitude deviation parameters, $\delta h_{\ell m}$, and constrain them directly from the data~\citep{Puecher:2022sfm,Gupta:2025xxx};  
    \item apply \bosch~\citep{Chiaramello:2025bhi}, which perturbs selected \ac{nr}-calibrated parameters in \TEOBResumSDali's dynamics and waveform generation.
\end{inparaenum}
This second method is conceptually similar to the parameterised-deviation scheme in the \texttt{pSEOBNR} framework~\citep{Brito:2018rfr, Ghosh:2021mrv, Maggio:2022hre, Pompili:2025cdc},
used in \ac{lvk} analyses of \event\ and earlier events~\citep{LIGOScientific:2020tif, LIGOScientific:2021sio} to constrain deviations in \ac{qnm} frequencies and test the \ac{imr}-faithfulness of the ringdown within \ac{gr}.

In our analysis, \bosch\ focuses on three key parameters: $\delta M_f$ and $\delta\chi_f$, which modify the \ac{nr}-informed remnant mass and spin, and $\delta A_{22}^{\mathrm{peak}}$, which alters
the peak amplitude of the dominant $(2,\pm2)$ mode. Changes in $\delta M_f$ and $\delta\chi_f$ affect the complex \ac{qnm} frequencies of $h_{\ell m}^{\mathrm{merger-RD}}(t)$,
leading to measurable shifts in the ringdown phase and damping times, while $\delta A_{22}^{\mathrm{peak}}$ perturbs the waveform amplitude near merger. Full details of the \bosch\ implementation
are given in~\citet{Chiaramello:2025bhi}.

Because \TEOBResumSDali\ models the inspiral–plunge with eccentric \ac{bbh} dynamics refined by \ac{nr} data,
\bosch\ can also quantify deviations from eccentric late-inspiral dynamics. However, \TEOBResumSDali\ currently transitions to a non-eccentric prescription for the merger–ringdown, assuming that binaries circularise sufficiently
by this stage—a simplification we plan to address in future work.\footnote{This limitation does not affect the results presented here, as we show in the following.}
As a result, \bosch\ can measure deviations from a non-eccentric post-merger baseline. \\

\begin{figure}
    \centering
    \includegraphics[width=0.49\textwidth]{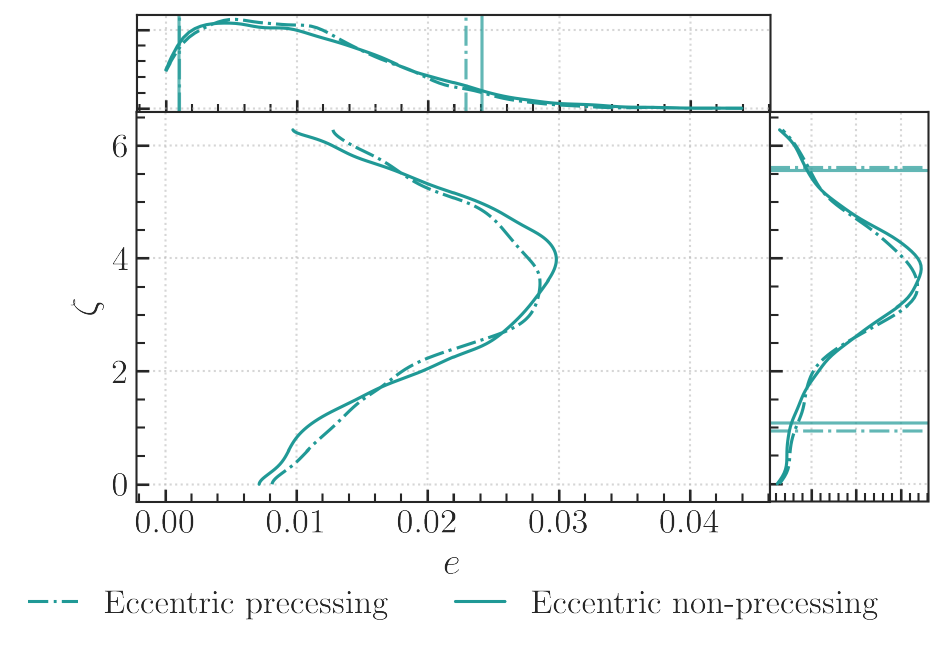}
    \caption{Posterior distributions for orbital eccentricity $e$ and the waveform parameter $\zeta$ obtained with eccentric precessing (dash-dotted) and eccentric non-precessing (solid) models. Both analyses give consistent results, with median values $e \simeq 0.01$ and $\zeta \simeq 3.5$ for both cases.}
    \label{fig:eccentricy-mean-anomaly}
\end{figure}

\noindent{\textbf{Astrophysical nature of GW250114's source:}}
The \ac{lvk} discovery and companion analyses established that \event\ originated from a quasi-circular, low-spin \ac{bbh} with negligible eccentricity~\citep{LIGOScientific:2025epi, LIGOScientific:2025obp}.
Here we re-examine this astrophysical picture using an extended set of waveform hypotheses, to independently validate and quantify the \ac{lvk} conclusions, and assess the
robustness of the source characterisation across a broader modelling space. To do this, we analyse 8\,s of publicly available Advanced LIGO data around \event\ using four different \ac{bbh} source hypothesis (see the \ac{sm} for details)~\citep{LIGOScientific:2025snk}:  
\begin{inparaenum}[(1)]
    \item non-eccentric, aligned-spin;  
    \item non-eccentric, precessing-spin;  
    \item eccentric, aligned-spin;  
    \item eccentric, precessing-spin.
\end{inparaenum}

Across all cases, the data favour a quasi-spherical, i.e., a non-eccentric but precessing configuration, consistent with the \ac{lvk} findings, which didn't report statistical evidence.  
Allowing for spin precession yields a slightly higher likelihood; removing precession reduces the maximum likelihood by $\Delta\max\ln\mathcal{L}\simeq -0.6$ and the Bayesian evidence by $\Delta\log_{10}\mathcal{B} = -0.12 \pm 0.11$.  
Eccentric models---whether aligned-spin or precessing---perform worse, with $\Delta\max\ln\mathcal{L}\simeq -1$ and $\Delta\log_{10}\mathcal{B} \lesssim -1.55 \pm 0.12$, and are thus more strongly disfavoured.
Our eccentricity constraints also closely match those obtained by the \ac{lvk}~\citep{LIGOScientific:2025epi, LIGOScientific:2025obp}. Figure~\ref{fig:eccentricy-mean-anomaly} shows the joint posterior on orbital eccentricity $e$ and mean anomaly $\zeta$ obtained using uniform priors $e\in(0,0.4)$ and $\zeta\in(0,2\pi)$. We obtain 
\( e = 0.01^{+0.01}_{-0.01}\). Together with the Bayes factors quoted above, these results strengthen \ac{lvk}'s claim that GW250114 is non-eccentric.
\begin{figure}
    \centering
    \includegraphics[width=0.49\textwidth]{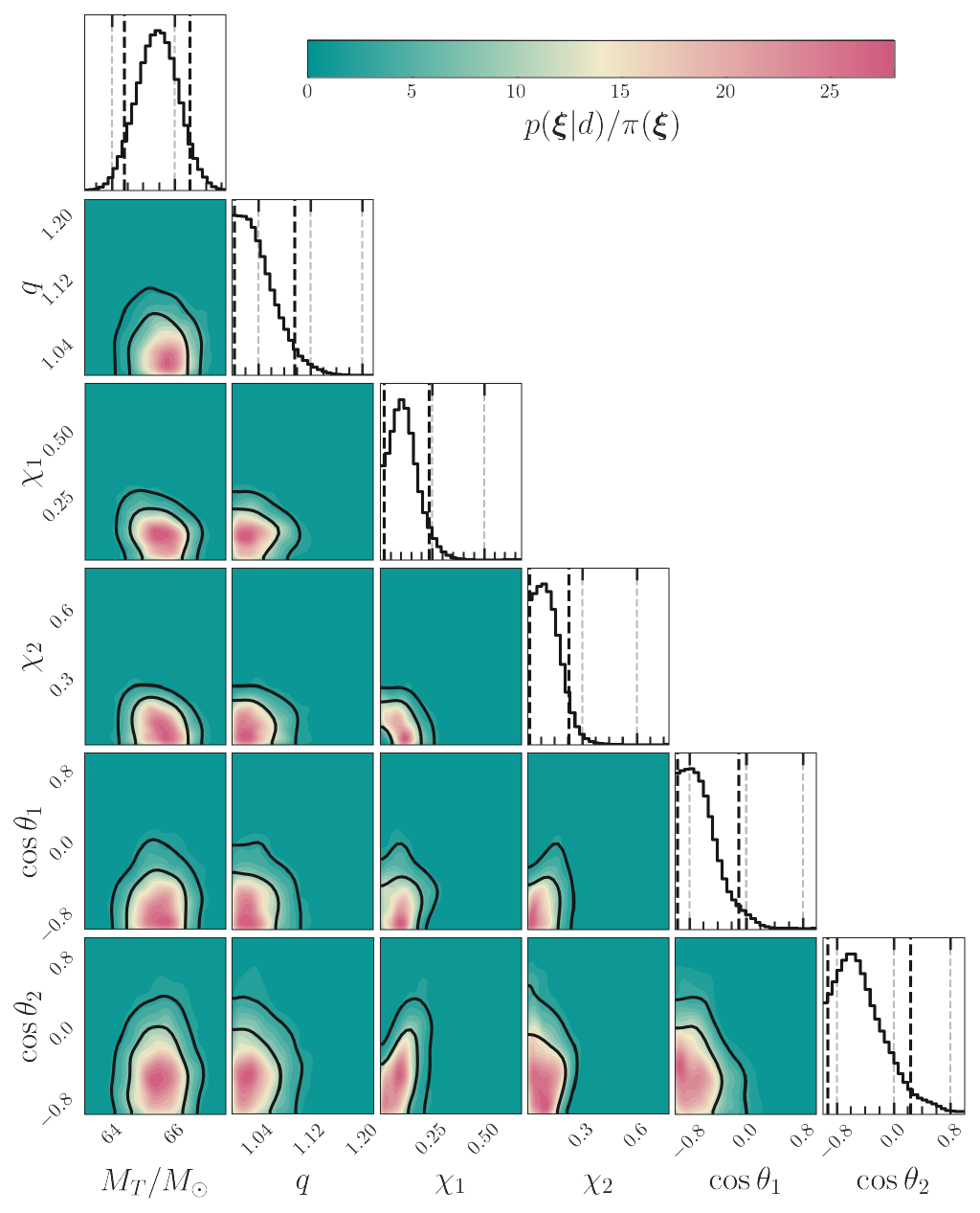}
    \caption{
    Posterior distributions of mass and spin parameters for GW250114. The 2D panels show the posteriors colored by the local posterior-to-prior density ratio $p(\boldsymbol{\xi}|d)/\pi(\boldsymbol{\xi})$, with warmer colors indicating greater support from the data. Black contours enclose 68\% and 90\% credible regions. The data favor an approximately equal-mass binary ($q \sim 1$) with low spin magnitudes ($\chi_{1,2} \sim 0.1$) and misaligned spins, as evidenced by the broad distributions of tilt angle cosines $\cos\theta_{1,2}$ measured relative to the orbital angular momentum.}
    \label{fig:gr-corner}
\end{figure}
Under the preferred (quasi-spherical) hypothesis, we recover source parameters consistent with the \ac{lvk}'s estimates. 
We infer a source-frame total mass of \(M_T = 65.7^{+0.8}_{-0.9}\,M_\odot\) and mass ratio \(q = 1.03^{+0.06}_{-0.03}\),
with low component spins, \(\chi_1 = 0.13^{+0.11}_{-0.11}\) and \(\chi_2 = 0.11^{+0.16}_{-0.09}\), as shown in Fig.~\ref{fig:gr-corner}.
These correspond to individual masses of \(m_1 = 33.3^{+0.9}_{-0.7}\,M_\odot\) and \(m_2 = 32.2^{+0.7}_{-1.0}\,M_\odot\).
Based on these estimates and data-driven \ac{gw}-based population inference results~\citep{KAGRA:2021duu}, we predict that the progenitors are likely first-generation
(1G) \acp{bh} with \(\geq 96\%\) certainty (see the \ac{sm} for details).

The posteriors in \(\cos\theta_{1,2}\) translate into spin-orbit tilt angles, defined with respect to the orbital angular momentum,
of \(\theta_{1} \in (105^{\circ},155^{\circ})\) and \(\theta_{2} \in (92^{\circ},138^{\circ})\) at local posterior-prior density ratio
\(p(\boldsymbol{\xi}|d)/\pi(\boldsymbol{\xi}) > 15\), thereby corroborating the signal's quasi-spherical \ac{bbh} origin.
These results are broadly consistent with \ac{lvk}'s findings obtained using the \texttt{NRSur7dq4} waveform model (see the \ac{sm} for details).  

From the posteriors and \ac{nr}-informed fits~\citep{Hofmann:2016yih, Jimenez-Forteza:2016oae}, we predict that the merger remnant has a final mass
$M_f = 63^{+1}_{-1}\,M_\odot\ $ and spin $\chi_f = 0.67^{+0.01}_{-0.01}$. Using the surrogate fits of \citet{Varma:2019csw, Varma:2020nbm},
whose consistency with \TEOBResumSDali\ we have explicitly verified across the posterior-supported region (see the \ac{sm} for details),
we estimate that the remnant's recoil speed is \(v_\mathrm{kick} = 141_{-117}^{+324}\,\mathrm{km/s}\).
The remnant's mass places it within the predicted pair-instability gap of \(60-130\,M_\odot\)~\citep{Woosley:2021xba, Farmer:2019jed},
and depending on the host environment, it may have either been ejected from or retained by it~\citep{Gerosa:2021mno}. \\

\begin{figure}
    \centering
    \includegraphics[width=0.49\textwidth]{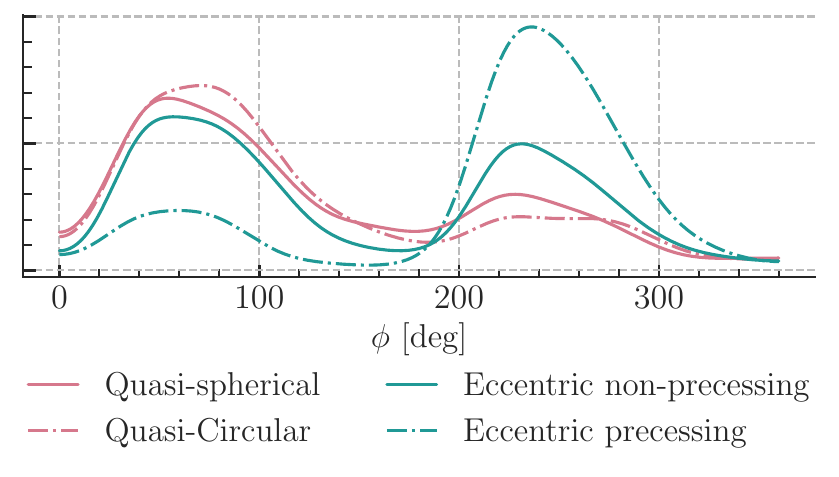}
    \caption{Posterior of azimuthal angle (measured in degrees at a reference frequency of 20\,Hz) for various \ac{bbh} models. The data favors a source with a bimodal azimuth separated by \(\pi\), reflecting the binary's symmetry under rotation \(\phi \rightarrow \phi + \pi\).}
    \label{fig:azimuth}
\end{figure}

\noindent{\textbf{Detectability and the temporal structure of the $(4, \pm 4)$ modes:}}
The detectability of higher-order gravitational-wave multipoles depends on the source's properties and orientation. 
For \event, the source's near mass symmetry and negligible spins suppress the odd-$m$ modes, and the radiation is dominated by the $(2,\pm2)$ component with no significant power mixing between modes of fixed $\ell$. 
Accordingly, based on the expected hierarchy of mode amplitudes, the $(\ell,m) = (4,\pm4)$ mode is likely to be the strongest contributor after the quadrupole.
This expectation is supported by the bimodal posterior in the azimuthal angle $\phi$, separated by $\pi$, reflecting the binary's inherent $\phi \rightarrow \phi + \pi$ symmetry (see Fig.~\ref{fig:azimuth}).

To ascertain the presence of the $(4,\pm4)$ modes, we compare the full–multipolar \TEOBResumSDali\ model with an otherwise identical version excluding these harmonics. The run including $(4,\pm4)$ yields a higher maximum log-likelihood, with $\Delta \ln \mathcal{L}_{\max} \simeq 11.5$, and a modest Bayes factor $\log_{10} \mathcal{B} \simeq 0.66$, 
with the maximum \ac{snr} gain ($\approx 4.79$) being comparable to the ``optimal'' \ac{snr} upper limit obtained for these modes in \citet{LIGOScientific:2025epi} using the formalism of \citet{Mills:2020thr}.
The modest Bayes factor, despite the substantial log-likelihood increase, reflects that the higher-harmonic improvement is restricted to a narrow region of parameter space and is therefore penalised by model complexity.\\

\begin{figure}
    \centering
    \includegraphics[width=0.49\textwidth]{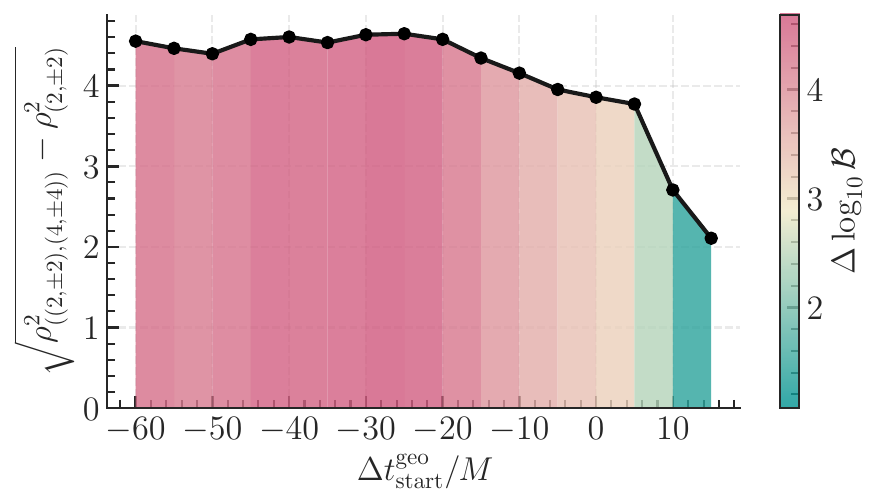}
    \caption{Time-gated analysis of GW250114 to test the contribution of the \((4,\pm4)\) mode as a function of the start time \(\Delta t^\mathrm{geo}_\mathrm{start}/M\), measured relative to the signal peak in units of remnant black-hole mass. The curves show the change in the log Bayes factor \(\Delta \log_{10} \mathcal{B}\) (colored) and the improvement in matched-filter \ac{snr} relative to a quadrupole-only model (black). The \((4,\pm4)\) contribution remains detectable even post signal's peak, thereby showing that GW250114 is the first \ac{gw} event with a measurable \((4,\pm4)\) contribution persisting into the post-merger phase.}
    \label{fig:temporal}
\end{figure}

We strengthen this finding by identifying when this multipole makes the most significant contribution to the signal.
To this end, we use a time-domain likelihood (see \ac{sm}), progressively excluding data before a start time \(\Delta t^\mathrm{geo}_\mathrm{start}\)~\citep{Chandra:2025ipu},
defined relative to the signal peak in units of the remnant black-hole mass. At each time step, we compare the \TEOBResumSDali\ model in its quasi-spherical configuration, including both the quadrupole and \((4,\pm4)\) multipole, against its quadrupole-only analogue.
We fix the sky location and peak time \(t_\mathrm{peak}^\mathrm{geo}\) to their maximum-likelihood values for this analysis.
As shown in Fig.~\ref{fig:temporal}, the \((4,\pm4)\) multipole has a detectable imprint even in the post-peak portion of the signal, up to \(\Delta t^\mathrm{geo}_\mathrm{start}/M = 5\), beyond which the maximum matched-filter \ac{snr} relative to a quadrupole-only model drops below 3.

Moreover, as the plot shows, restricting to late-time data yields $\log_{10}\mathcal{B} \sim 1-4$ against the quadrupole-only
hypothesis, larger than the one quoted from the full inspiral-merger-ringdown analysis. This is because in a full signal analysis, the earlier inspiral dilutes the relative information carried by the \((4,\pm4)\) mode. These values are also larger than those obtained from the \texttt{KerrPostmerger} analysis of \citet{LIGOScientific:2025epi}; we attribute this difference primarily to prior choices and model features. \footnote{\texttt{KerrPostmerger} samples over each mode's phase at merger independently, whereas in our framework these phases are fixed by the inspiral–plunge dynamics~\citep{Gennari:2023gmx}. Additionally, \texttt{KerrPostmerger} employs a broader mass prior.}\footnote{\texttt{pSEOBNR} analyses indirectly suggest such a contribution by constraining deviations in the 440 \ac{qnm} frequencies, and by comparing the results with simulated signals in which the $(4,\pm4)$ multipole is removed, although—as quoted in the paper—``the analysis cannot by itself establish whether the 440 QNM is present in the data''.} \\

\begin{figure}
    \centering
    \includegraphics[width=0.49\textwidth]{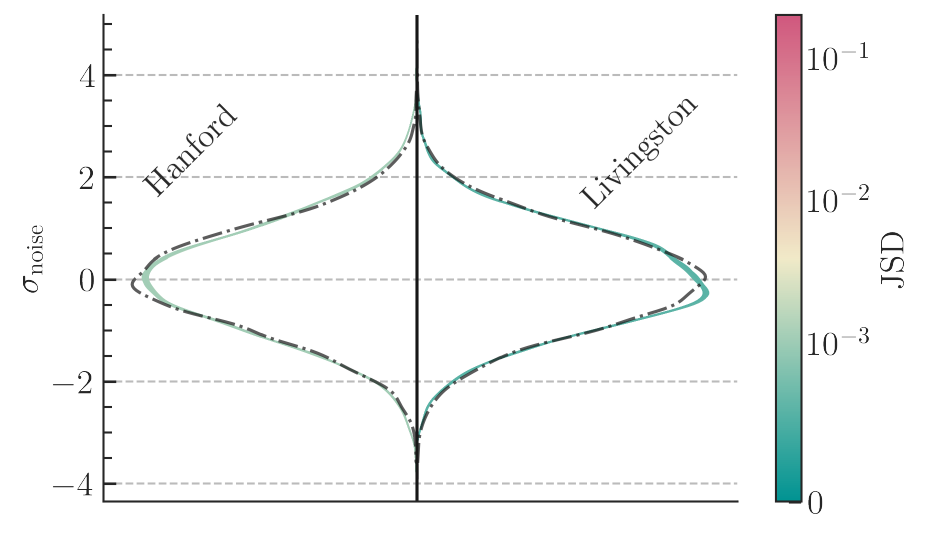}
    \caption{Residual distributions for Hanford (left) and Livingston (right) detectors after subtracting the whitened \TEOBResumSDali~posterior waveforms from whitened strain data. Individual curves show posterior samples, colored by their Jensen–Shannon divergence from the expected white noise distribution (black dashed). We find that the residuals are statistically indistinguishable from noise ($\text{JSD} = 0.008 \pm 3\times10^{-4}$, $p > 0.05$ for all samples), thereby indicating that the data is well described by \ac{gr}-consistent \ac{bbh} waveforms.}
    \label{fig:residuals}
\end{figure}

\begin{figure}
    \centering
    \includegraphics[width=0.49\textwidth]{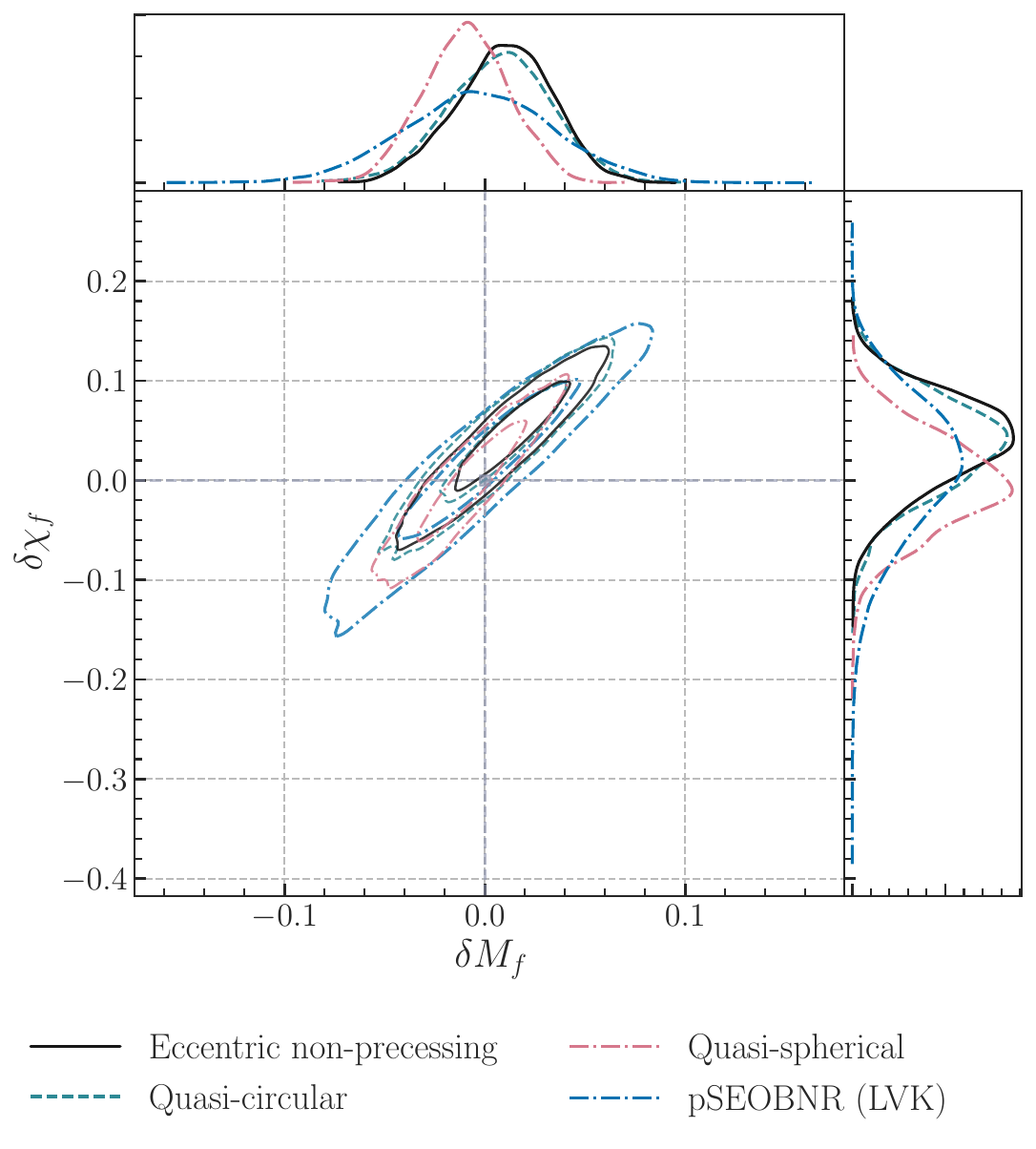}
    \caption{Posterior for the fractional deviations in the final mass, $\delta M_f$, and final spin, $\delta \chi_f$, relative to the \ac{gr} predictions for the quasi-spherical, quasi-circular, and eccentric, non-precessing model. The outer contours enclose the \(90\%\) credible regions and in both non-eccentric cases the \ac{gr} expectation $\delta M_f = \delta \chi_f = 0$ lies within 50\% credible level. For comparison, we also show the corresponding constraints derived from the publicly available \texttt{pSEOBNR} (LVK) samples (dashed dotted blue). To enable a direct comparison, we first map the \texttt{pSEOBNR} \ac{bbh} posteriors to obtain the \ac{imr}-predicted quantities, and separately map the deviated fundamental mode \ac{qnm} spectra samples to their corresponding remnant properties. From these, we compute the fractional deviations. As shown, the \texttt{pSEOBNR} posteriors also include the \ac{gr} value at their \(50\%\) credible interval. Our constraints are tighter, reflecting the additional information contributed by the subdominant angular mode present in our analysis.}
    \label{fig:deltaMf_chif}
\end{figure}

\noindent{\textbf{Signal's consistency with GR:}} 
We begin by assessing the consistency of the data with the \ac{gr} prediction using a modified residual analysis. The \ac{lvk} companion paper performed this test by subtracting the
\emph{maximum-likelihood} waveform from the data and showing that the residual network \ac{snr} is compatible with the noise background with a Kolmogorov–Smirnov p-value of 0.34.
Here we complement that study by performing a slightly different test: instead of subtracting a single waveform, we subtract the \emph{full ensemble}
of posterior-supported quasi-spherical \TEOBResumSDali\ waveforms from the data. This approach avoids relying on the single maximum-likelihood waveform returned by an
off-the-shelf stochastic sampler, which is not optimised to locate the true maximum log-likelihood point in high-dimensional parameter space~\citep{Speagle:2019ivv}.

Figure~\ref{fig:residuals} shows the resulting whitened residuals, obtained by subtracting the whitened posterior waveforms from the whitened strain and comparing the
distribution with a zero-mean univariate Gaussian. Across the posterior ensemble, the residuals are statistically indistinguishable from the expected coloured Gaussian noise.
We measure Jensen–Shannon divergences of $0.088 \pm 0.003$ and Kolmogorov-Smirnov $p$-values of $0.23 \pm 0.24$, all satisfying $\mathrm{JSD} < 8 \times 10^{-3}$ and $p > 0.05$.
These \textit{show that the general-relativistic \TEOBResumSDali\ waveform model fully captures the gravitational-wave signal} and the remaining fluctuations are consistent with detector noise. \\

\begin{figure}
    \centering
    \includegraphics[width=0.49\textwidth]{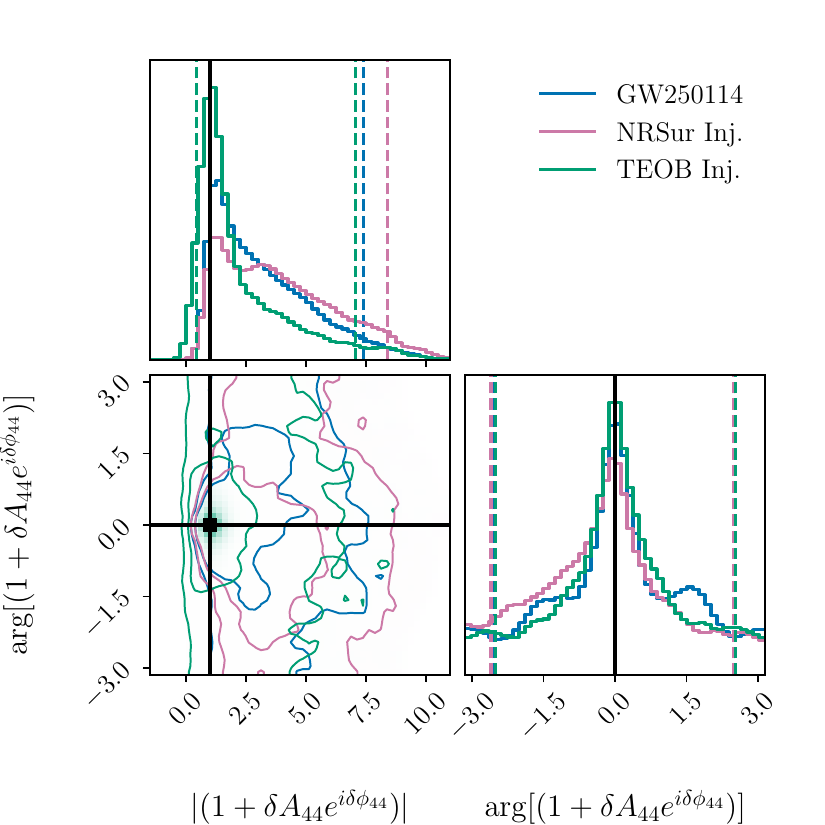}
    \caption{Posterior distributions of the fractional $(4,4)$-mode deviation,  $\delta h = (1 + \delta A_{44} e^{i \delta \phi_{44}})$. 
    The blue curve shows the observed event, the pink curve a \texttt{NRSur7dq4} synthetic signal recovered with \TEOBResumSDali\ in zero noise, and the green curve a
    matched \TEOBResumSDali--\TEOBResumSDali\ recovery. 
    In all cases, the $(4,4)$ mode deviation is consistent with \ac{gr}, yielding a posterior peaked near the value of $1$ (vertical black line).
    However, in the event and the mismatched synthetic signal, we observe a slight shift of the posterior amplitude to the right,
    likely due to waveform systematics.}
    \label{fig:amplitude-deviations}
\end{figure}

Assuming that the dominant quadrupole $(\ell,m) = (2,\pm2)$ mode agrees with \ac{gr}, we next test the subdominant $(4,\pm4)$ multipole under a
quasi-circular signal hypothesis.~\footnote{As shown earlier, precessing and non-precessing configurations are statistically indistinguishable for this event, supporting the use of the simpler non-precessing analysis.
Also, precession leads to significant power mixing between modes of fixed \(\ell\),
indicating one needs to add these deviations in the co-precessing frame, which we don't.} 
Following and extending~\citet{Puecher:2022sfm, Gupta:2025xxx}, we introduce a \emph{time-independent fractional mode deviation}, parameterised by an
amplitude and a phase deviation via \( h_{44} \ \rightarrow\ h_{44}\,(1+\delta A_{44} e^{i \delta \phi_{44}})\),
and constrain it directly from the data while enforcing $h_{\ell,-m} = (-1)^\ell h_{\ell m}^\ast$.
Figure~\ref{fig:amplitude-deviations} shows the deviations posteriors, which are 
consistent with the \ac{gr} prediction of \( (1+\delta A_{44} e^{i \delta \phi_{44}}) = 1 \)
at the 90\% credible level, though the amplitude posterior shows a long tail towards higher values
and a shift of the peak away from the \ac{gr} expectation.
To test the robustness of this finding, we perform two additional zero-noise
simulations with a \event-like signal:
\begin{inparaenum}[(i)]
    \item matched \TEOBResumSDali--\TEOBResumSDali\ recovery, and
    \item unmatched \texttt{NRSur7dq4}--\TEOBResumSDali\ recovery.
\end{inparaenum}
The former yields a unimodal posterior peaked at the \ac{gr} value, while the latter reproduces the shift in
in peak amplitude deviation similar to that observed in the event analysis.
This suggests that waveform systematics are responsible for this feature.
\emph{To our knowledge, this is one of the first attempts to directly test the subdominant $(4,\pm4)$ multipole's
amplitude consistency with \ac{gr} in a \ac{bbh} signal.} \\

Next, we perform two complementary parameterised tests of the plunge--merger--ringdown regime.  
Our first test introduces fractional deviations in the remnant mass and spin, $(\delta M_f, \delta\chi_f)$, relative to their \ac{gr} values, sampling both from uniform priors over $(-0.2,0.2)$.  
Figure~\ref{fig:deltaMf_chif} shows the resulting posteriors for three non-precessing signal hypotheses: quasi-spherical (dash-dotted rose), quasi-circular (dashed teal), and eccentric (solid black).  
In both non-eccentric cases, the \ac{gr} point $(\delta M_f=\delta\chi_f=0)$ lies well within the $\sim 50\%$ credible region while for the eccentric case it lies within the \(70\%\) credible region.  
The corresponding Savage--Dickey ratios,
$\log_{10}[p(\delta M_f,\delta\chi_f|d)/\pi(\delta M_f,\delta\chi_f)] = 1.51 \mathrm{~and~} 2.84$,
provide positive support for \ac{gr}-consistent post-inspiral dynamics for both non-eccentric signal source hypothesis.

Additionally, we evaluate the Bayesian support for \ac{gr}-consistent plunge--merger--ringdown against beyond-\ac{gr} extensions. For quasi-circular binaries, we obtain $\Delta\log_{10}\mathcal{B}^{\rm GR}_{\rm bGR} = 2.30 \pm 0.12$ ($\mathcal{B}\simeq200$), constituting decisive evidence for \ac{gr}. Although the maximum likelihood improves only marginally ($\Delta\ln\mathcal{L}_{\rm max}=0.49$), the deviation models are penalised by their larger prior volume. Quasi-spherical binaries show similar behaviour, with $\Delta\log_{10}\mathcal{B} = 2.63 \pm 0.12$ and $\Delta\ln\mathcal{L}_{\rm max}=0.57$, again strongly favouring \ac{gr}. Taken together, both the metrics: the Savage-Dickey ratios and the Bayes factors indicate no evidence for a \ac{gr}-inconsistent \ac{bbh} plunge–merger–ringdown.

As previously mentioned, this analysis, though similar in spirit to the \texttt{pSEOBNR} test performed by the \ac{lvk}~\citep{Brito:2018rfr, Pompili:2025cdc, LIGOScientific:2025obp},
differs in its implementation. \texttt{pSEOBNR} constrains deviations from the \ac{imr}-expected \ac{qnm} frequencies, while we directly
constrain deviations in the \ac{imr}-expected remnant properties that inform the \ac{qnm} frequencies that is used for modelling the post-peak waveform.
{Therefore, to compare the results, we use the \ac{lvk}'s publicly available \texttt{pSEOBNR} posterior samples and reconstruct the corresponding remnant properties. Starting from the intrinsic binary parameters, we use \ac{nr} fitting formulae to obtain the \ac{gr}-consistent inspiral predictions for the remnant, \(\bigl(M_f^{\rm IMR}(1+z),\,\chi_f^{\rm IMR}\bigr)\).  
Independently, we map the measured quasi-normal–mode quantities \(\bigl(f_{220}(1+\delta f_{220}),\,\tau_{220}(1+\delta\tau_{220})\bigr)\) to the corresponding remnant mass and spin, \(\bigl(M_f^{\rm dev}(1+z),\,\chi_f^{\rm dev}\bigr)\), using the relations of~\citet{Berti:2006wq}.  
We then quantify departures from \ac{gr} through \(
\delta X_f \equiv X_f^{\rm dev}/X_f^{\rm IMR} - 1,
\)
for \(X_f\in\{M_f,\chi_f\}\).
As shown in Fig.~\ref{fig:deltaMf_chif}, the resulting \texttt{pSEOBNR} constraints include the \ac{gr} prediction also within the \(50\%\) credible interval. Our bounds are somewhat tighter, consistent with the fact that our analysis leverages information from the subdominant \((4,\pm4)\) multipole in addition to the dominant mode.}

\begin{figure}[]
    \centering
    \includegraphics[width=0.49\textwidth]{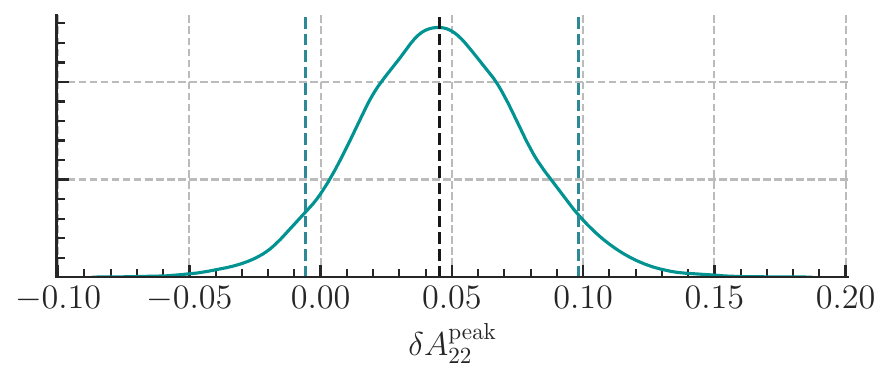}
    \caption{Posterior distribution of the fractional deviation in the peak amplitude of the dominant $(2,2)$ mode, $\delta A_{22}^{\rm peak}$. The outer dashed blue contour encloses the 90\% credible region, while the dashed black line marks the median. The \ac{gr} expectation, $\delta A_{22}^{\rm peak}=0$, lies at the 82\% credible level, indicating consistency with \ac{gr}.}
    \label{fig:merger-amplitude-deviations}
\end{figure}

Our second test allows a fractional deviation in the peak amplitude of the dominant $(2,2)$ mode, $\delta A_{22}^{\rm peak}$, sampling from a uniform prior on $(-1, 4)$ under the quasi-circular assumption. For such an assumption, symmetry imposes the relation $h_{\ell,-m} = (-1)^{\ell} h_{\ell,m}^{\ast}$, ensuring that the $(2,-2)$ mode is fully determined by the $(2,2)$ mode and removing additional freedom in the waveform morphology. We find that $\delta A_{22}^{\rm peak} = 0$ lies within the $\sim$82\% credible interval, consistent with \ac{gr} (See Fig.~\ref{fig:merger-amplitude-deviations}) and has a posterior width of $\mathcal{O}(10^{-1})$. For comparison, this value is just one to two orders of magnitude larger than the typical systematic uncertainty in \ac{nr} simulations obtained from different codes (see e.g. Fig.~8 of~\citet{Rashti:2024yoc}). This fact is rather remarkable, showcasing how \ac{gw} observations are soon to enter the high precision regime. \\

\noindent{\textbf{Discussion:}}
\label{sec:discussion}
Our complementary analysis confirms and strengthens~\citet{LIGOScientific:2025epi, LIGOScientific:2025obp} results, indicating
\event\ as a nearly mass-symmetric quasi-spherical \ac{bbh} merger with total mass \(M_T \simeq 66\,M_\odot\) and low spins \(\chi_{1,2} \simeq 0.1\),
decisively disfavouring eccentric configurations (\(\mathcal{B}\sim20\)).
These properties point to a first-generation binary formed through isolated field evolution~\citep{Postnov:2014tza} with $\geq 96\%$ probability.
Additionally owing to these properties and the signal's high \ac{snr} (\(\gtrsim75\)), we obtain decisive evidence for the subdominant \((4,\pm4)\) multipoles,
detectable up to \(\Delta t^\mathrm{geo}_\mathrm{start}=5M\) beyond the signal's peak, thereby strengthening the finding of \citet{LIGOScientific:2025obp}.
 
We then subject \ac{gr} to a trio of stringent, complementary tests in the strong-field, non-linear regime with this signal. 
First, we perform a residual analysis, subtracting the full ensemble of posterior-supported waveforms from the data, and confirm that \TEOBResumSDali~fully captures the signal,
with no statistically significant deviations from noise ($p \simeq 0.3$). 
Second, we carry out the one of the first direct parameterised test of the subdominant $(4,\pm4)$ multipole's amplitude consistency with \ac{gr}$\,$ in a \ac{bbh} signal,
introducing a fractional mode deviation and recovering its posterior from the data. We find that the \ac{gr} value lies within the 90\% credible
region, but the deviation posterior exhibits a mild shift in amplitude, a structure reproduced in simulated signal analysis where a \texttt{NRSur7dq4} waveform is
recovered using \TEOBResumSDali, but not in matched \TEOBResumSDali--\TEOBResumSDali\ recoveries. The agreement between the signal and mismatched simulated signals
points to waveform systematics in the $(4,\pm4)$ mode of \TEOBResumSDali.
Lastly, using the first parameterised eccentric \ac{bbh} waveform model, we find a \ac{gr}-consistent plunge--merger--ringdown portion, with the \ac{gr} expectation
$(\delta M_f = \delta \chi_f = 0)$ lying well within the 50\% credible region for both non-eccentric cases and \(\delta A^\mathrm{peak}_{22}\) lying inside the 90\% credible region.
Regarding the latter, we note that while finalising the paper, we became aware of \citet{Grimaldi2025} that also constrains the GW250114 merger amplitude. \\

Collectively these analyses, together with those in \citet{LIGOScientific:2025epi, LIGOScientific:2025obp}, provide compelling and multifaceted evidence for the
validity of \ac{gr} in the dynamical, strong-field gravity of a \ac{bbh} coalescence. \event\ thus stands as a watershed event, heralding a new era of precision
strong-gravity physics where future high-\ac{snr} detections will enable even more stringent constraints on potential deviations from \ac{gr}. \\

\noindent{\textbf{Acknowledgements:}} The authors thank Arnab Dhani, Ish Gupta, B. Sathyaprakash, and Juan Calderón Bustillo
for their comments and valuable suggestions, which greatly improved this paper. 
They also appreciate Alessandra Buoananno's and Elisa Maggio's comments and constructive feedback. 
RG also thanks K.~Glenn for help in cleaning the manuscript, while KC thanks A. Hussain.
KC acknowledges support from NSF grants AST-2307147, PHY-2207638, and  PHY-2308886.
RG acknowledges support from NSF Grant PHY-2020275 (Network for Neutrinos, Nuclear Astrophysics, and Symmetries (N3AS))
This work has made use of data obtained from the Gravitational Wave Open Science Center (\href{gwosc.org}{gwosc.org}), a service of the LIGO Scientific Collaboration, the Virgo Collaboration, and KAGRA. This material is based upon work supported by NSF's LIGO Laboratory which is a major facility fully funded by the National Science Foundation, as well as the Science and Technology Facilities Council (STFC) of the United Kingdom, the Max-Planck-Society (MPS), and the State of Niedersachsen/Germany for support of the construction of Advanced LIGO and construction and operation of the GEO600 detector. Additional support for Advanced LIGO was provided by the Australian Research Council.
Finally, the authors acknowledge using the Nemo cluster at the University of Wisconsin-Milwaukee (supported by NSF grants PHY-1626190 and PHY-2110594) and the Gwave cluster at Pennsylvania State University (supported by NSF grants OAC-2346596, OAC-2201445, OAC-2103662, OAC-2018299, and PHY-2110594) for the computational analyses presented here.

\bibliography{references}{}

@article{LIGOScientific:2025epi,
    author = "Abac, A. G. and others",
    collaboration = "LIGO Scientific, KAGRA, Virgo",
    title = "{GW250114: Testing Hawking{\textquoteright}s Area Law and the Kerr Nature of Black Holes}",
    eprint = "2509.08054",
    archivePrefix = "arXiv",
    primaryClass = "gr-qc",
    reportNumber = "LIGO-P2500421",
    doi = "10.1103/kw5g-d732",
    journal = "Phys. Rev. Lett.",
    volume = "135",
    number = "11",
    pages = "111403",
    year = "2025"
}

@article{LIGOScientific:2025obp,
    author = "Abac, A. G. and others",
    collaboration = "LIGO Scientific, VIRGO, KAGRA",
    title = "{Black Hole Spectroscopy and Tests of General Relativity with GW250114}",
    eprint = "2509.08099",
    archivePrefix = "arXiv",
    primaryClass = "gr-qc",
    reportNumber = "LIGO P2500461",
    month = "9",
    year = "2025"
}

@article{Romero-Shaw:2020owr,
    author = "Romero-Shaw, I. M. and others",
    title = "{Bayesian inference for compact binary coalescences with bilby: validation and application to the first LIGO{\textendash}Virgo gravitational-wave transient catalogue}",
    eprint = "2006.00714",
    archivePrefix = "arXiv",
    primaryClass = "astro-ph.IM",
    doi = "10.1093/mnras/staa2850",
    journal = "Mon. Not. Roy. Astron. Soc.",
    volume = "499",
    number = "3",
    pages = "3295--3319",
    year = "2020"
}

@article{Speagle:2019ivv,
    author = "Speagle, Joshua S.",
    title = "{dynesty: a dynamic nested sampling package for estimating Bayesian posteriors and evidences}",
    eprint = "1904.02180",
    archivePrefix = "arXiv",
    primaryClass = "astro-ph.IM",
    doi = "10.1093/mnras/staa278",
    journal = "Mon. Not. Roy. Astron. Soc.",
    volume = "493",
    number = "3",
    pages = "3132--3158",
    year = "2020"
}

@article{Ashton:2018jfp,
    author = "Ashton, Gregory and others",
    title = "{BILBY: A user-friendly Bayesian inference library for gravitational-wave astronomy}",
    eprint = "1811.02042",
    archivePrefix = "arXiv",
    primaryClass = "astro-ph.IM",
    doi = "10.3847/1538-4365/ab06fc",
    journal = "Astrophys. J. Suppl.",
    volume = "241",
    number = "2",
    pages = "27",
    year = "2019"
}

@article{Puecher:2022sfm,
    author = "Puecher, Anna and Kalaghatgi, Chinmay and Roy, Soumen and Setyawati, Yoshinta and Gupta, Ish and Sathyaprakash, B. S. and Van Den Broeck, Chris",
    title = "{Testing general relativity using higher-order modes of gravitational waves from binary black holes}",
    eprint = "2205.09062",
    archivePrefix = "arXiv",
    primaryClass = "gr-qc",
    reportNumber = "LIGO DCC: LIGO-P2200140",
    doi = "10.1103/PhysRevD.106.082003",
    journal = "Phys. Rev. D",
    volume = "106",
    number = "8",
    pages = "082003",
    year = "2022"
}

@article{DelPozzo:2016kmd,
    author = "Del Pozzo, Walter and Nagar, Alessandro",
    title = "{Analytic family of post-merger template waveforms}",
    eprint = "1606.03952",
    archivePrefix = "arXiv",
    primaryClass = "gr-qc",
    doi = "10.1103/PhysRevD.95.124034",
    journal = "Phys. Rev. D",
    volume = "95",
    number = "12",
    pages = "124034",
    year = "2017"
}

@article{Gamba:2024cvy,
    author = "Gamba, Rossella and Chiaramello, Danilo and Neogi, Sayan",
    title = "{Toward efficient effective-one-body models for generic, nonplanar orbits}",
    eprint = "2404.15408",
    archivePrefix = "arXiv",
    primaryClass = "gr-qc",
    doi = "10.1103/PhysRevD.110.024031",
    journal = "Phys. Rev. D",
    volume = "110",
    number = "2",
    pages = "024031",
    year = "2024"
}

@article{Damour:2014yha,
    author = "Damour, Thibault and Nagar, Alessandro",
    title = "{A new analytic representation of the ringdown waveform of coalescing spinning black hole binaries}",
    eprint = "1406.0401",
    archivePrefix = "arXiv",
    primaryClass = "gr-qc",
    doi = "10.1103/PhysRevD.90.024054",
    journal = "Phys. Rev. D",
    volume = "90",
    number = "2",
    pages = "024054",
    year = "2014"
}

@article{Albanesi:2025txj,
    author = "Albanesi, Simone and Gamba, Rossella and Bernuzzi, Sebastiano and Fontbut{\'e}, Joan and Gonzalez, Alejandra and Nagar, Alessandro",
    title = "{Effective-one-body modeling for generic compact binaries with arbitrary orbits}",
    eprint = "2503.14580",
    archivePrefix = "arXiv",
    primaryClass = "gr-qc",
    month = "3",
    year = "2025"
}

@article{Nagar:2024oyk,
    author = "Nagar, Alessandro and Chiaramello, Danilo and Gamba, Rossella and Albanesi, Simone and Bernuzzi, Sebastiano and Fantini, Veronica and Panzeri, Mattia and Rettegno, Piero",
    title = "{Effective-one-body waveform model for noncircularized, planar, coalescing black hole binaries. II. High accuracy by improving logarithmic terms in resummations}",
    eprint = "2407.04762",
    archivePrefix = "arXiv",
    primaryClass = "gr-qc",
    doi = "10.1103/PhysRevD.111.064050",
    journal = "Phys. Rev. D",
    volume = "111",
    number = "6",
    pages = "064050",
    year = "2025"
}

@article{Veitch:2009hd,
    author = "Veitch, J. and Vecchio, A.",
    title = "{Bayesian coherent analysis of in-spiral gravitational wave signals with a detector network}",
    eprint = "0911.3820",
    archivePrefix = "arXiv",
    primaryClass = "astro-ph.CO",
    reportNumber = "LIGO-P0900117",
    doi = "10.1103/PhysRevD.81.062003",
    journal = "Phys. Rev. D",
    volume = "81",
    pages = "062003",
    year = "2010"
}

@article{Finn:1992wt,
    author = "Finn, Lee S.",
    title = "{Detection, measurement and gravitational radiation}",
    eprint = "gr-qc/9209010",
    archivePrefix = "arXiv",
    reportNumber = "PRINT-93-0128 (NORTHWESTERN)",
    doi = "10.1103/PhysRevD.46.5236",
    journal = "Phys. Rev. D",
    volume = "46",
    pages = "5236--5249",
    year = "1992"
}

@article{Hofmann:2016yih,
    author = "Hofmann, Fabian and Barausse, Enrico and Rezzolla, Luciano",
    title = "{The final spin from binary black holes in quasi-circular orbits}",
    eprint = "1605.01938",
    archivePrefix = "arXiv",
    primaryClass = "gr-qc",
    doi = "10.3847/2041-8205/825/2/L19",
    journal = "Astrophys. J. Lett.",
    volume = "825",
    number = "2",
    pages = "L19",
    year = "2016"
}

@article{Jimenez-Forteza:2016oae,
    author = {Jim{\'e}nez-Forteza, Xisco and Keitel, David and Husa, Sascha and Hannam, Mark and Khan, Sebastian and P{\"u}rrer, Michael},
    title = "{Hierarchical data-driven approach to fitting numerical relativity data for nonprecessing binary black holes with an application to final spin and radiated energy}",
    eprint = "1611.00332",
    archivePrefix = "arXiv",
    primaryClass = "gr-qc",
    reportNumber = "LIGO-P1600270",
    doi = "10.1103/PhysRevD.95.064024",
    journal = "Phys. Rev. D",
    volume = "95",
    number = "6",
    pages = "064024",
    year = "2017"
}

@article{Chandramouli:2024vhw,
    author = "Chandramouli, Rohit S. and Prokup, Kaitlyn and Berti, Emanuele and Yunes, Nicol{\'a}s",
    title = "{Systematic biases due to waveform mismodeling in parametrized post-Einsteinian tests of general relativity: The impact of neglecting spin precession and higher modes}",
    eprint = "2410.06254",
    archivePrefix = "arXiv",
    primaryClass = "gr-qc",
    doi = "10.1103/PhysRevD.111.044026",
    journal = "Phys. Rev. D",
    volume = "111",
    number = "4",
    pages = "044026",
    year = "2025"
}

@article{Varma:2019csw,
    author = "Varma, Vijay and Field, Scott E. and Scheel, Mark A. and Blackman, Jonathan and Gerosa, Davide and Stein, Leo C. and Kidder, Lawrence E. and Pfeiffer, Harald P.",
    title = "{Surrogate models for precessing binary black hole simulations with unequal masses}",
    eprint = "1905.09300",
    archivePrefix = "arXiv",
    primaryClass = "gr-qc",
    doi = "10.1103/PhysRevResearch.1.033015",
    journal = "Phys. Rev. Research.",
    volume = "1",
    pages = "033015",
    year = "2019"
}

@article{Gupta:2024gun,
    author = "Gupta, Anuradha and others",
    title = "{Possible causes of false general relativity violations in gravitational wave observations}",
    eprint = "2405.02197",
    archivePrefix = "arXiv",
    primaryClass = "gr-qc",
    doi = "10.21468/SciPostPhysCommRep.5",
    month = "5",
    year = "2024"
}

@article{Maggio:2022hre,
    author = "Maggio, Elisa and Silva, Hector O. and Buonanno, Alessandra and Ghosh, Abhirup",
    title = "{Tests of general relativity in the nonlinear regime: A parametrized plunge-merger-ringdown gravitational waveform model}",
    eprint = "2212.09655",
    archivePrefix = "arXiv",
    primaryClass = "gr-qc",
    reportNumber = "LIGO-P2200377",
    doi = "10.1103/PhysRevD.108.024043",
    journal = "Phys. Rev. D",
    volume = "108",
    number = "2",
    pages = "024043",
    year = "2023"
}

@article{LIGOScientific:2025cmm,
    collaboration = "LIGO Scientific, VIRGO, Kagra",
    title = "{GW230814: investigation of a loud gravitational-wave signal observed with a single detector}",
    eprint = "2509.07348",
    archivePrefix = "arXiv",
    primaryClass = "gr-qc",
    reportNumber = "LIGO-P230814",
    month = "9",
    year = "2025"
}

@article{Hu:2022bji,
    author = "Hu, Qian and Veitch, John",
    title = "{Accumulating Errors in Tests of General Relativity with Gravitational Waves: Overlapping Signals and Inaccurate Waveforms}",
    eprint = "2210.04769",
    archivePrefix = "arXiv",
    primaryClass = "gr-qc",
    reportNumber = "ET-0211A-22",
    doi = "10.3847/1538-4357/acbc18",
    journal = "Astrophys. J.",
    volume = "945",
    number = "2",
    pages = "103",
    year = "2023"
}

@article{Varma:2020nbm,
    author = "Varma, Vijay and Isi, Maximiliano and Biscoveanu, Sylvia",
    title = "{Extracting the Gravitational Recoil from Black Hole Merger Signals}",
    eprint = "2002.00296",
    archivePrefix = "arXiv",
    primaryClass = "gr-qc",
    doi = "10.1103/PhysRevLett.124.101104",
    journal = "Phys. Rev. Lett.",
    volume = "124",
    number = "10",
    pages = "101104",
    year = "2020"
}

@article{Gerosa:2021mno,
    author = "Gerosa, Davide and Fishbach, Maya",
    title = "{Hierarchical mergers of stellar-mass black holes and their gravitational-wave signatures}",
    eprint = "2105.03439",
    archivePrefix = "arXiv",
    primaryClass = "astro-ph.HE",
    doi = "10.1038/s41550-021-01398-w",
    journal = "Nature Astron.",
    volume = "5",
    number = "8",
    pages = "749--760",
    year = "2021"
}

@article{Farmer:2019jed,
    author = "Farmer, R. and Renzo, M. and de Mink, S. E. and Marchant, P. and Justham, S.",
    title = "{Mind the gap: The location of the lower edge of the pair instability supernovae black hole mass gap}",
    eprint = "1910.12874",
    archivePrefix = "arXiv",
    primaryClass = "astro-ph.SR",
    doi = "10.3847/1538-4357/ab518b",
    month = "10",
    year = "2019"
}

@article{Woosley:2021xba,
    author = "Woosley, S. E. and Heger, Alexander",
    title = "{The Pair-Instability Mass Gap for Black Holes}",
    eprint = "2103.07933",
    archivePrefix = "arXiv",
    primaryClass = "astro-ph.SR",
    doi = "10.3847/2041-8213/abf2c4",
    journal = "Astrophys. J. Lett.",
    volume = "912",
    number = "2",
    pages = "L31",
    year = "2021"
}

@article{Krishnendu:2021fga,
    author = "Krishnendu, N. V. and Ohme, Frank",
    title = "{Testing General Relativity with Gravitational Waves: An Overview}",
    eprint = "2201.05418",
    archivePrefix = "arXiv",
    primaryClass = "gr-qc",
    doi = "10.3390/universe7120497",
    journal = "Universe",
    volume = "7",
    number = "12",
    pages = "497",
    year = "2021"
}

@article{Yunes:2025xwp,
    author = "Yunes, Nicol{\'a}s and Siemens, Xavier and Yagi, Kent",
    title = "{Gravitational-wave tests of general relativity with ground-based detectors and pulsar-timing arrays}",
    doi = "10.1007/s41114-024-00054-9",
    journal = "Living Rev. Rel.",
    volume = "28",
    number = "1",
    pages = "3",
    year = "2025"
}

@article{Brito:2018rfr,
    author = "Brito, Richard and Buonanno, Alessandra and Raymond, Vivien",
    title = "{Black-hole Spectroscopy by Making Full Use of Gravitational-Wave Modeling}",
    eprint = "1805.00293",
    archivePrefix = "arXiv",
    primaryClass = "gr-qc",
    doi = "10.1103/PhysRevD.98.084038",
    journal = "Phys. Rev. D",
    volume = "98",
    number = "8",
    pages = "084038",
    year = "2018"
}

@article{Buonanno:1998gg,
    author = "Buonanno, A. and Damour, T.",
    title = "{Effective one-body approach to general relativistic two-body dynamics}",
    eprint = "gr-qc/9811091",
    archivePrefix = "arXiv",
    reportNumber = "IHES-P-98-74",
    doi = "10.1103/PhysRevD.59.084006",
    journal = "Phys. Rev. D",
    volume = "59",
    pages = "084006",
    year = "1999"
}

@article{Buonanno:2000ef,
    author = "Buonanno, Alessandra and Damour, Thibault",
    title = "{Transition from inspiral to plunge in binary black hole coalescences}",
    eprint = "gr-qc/0001013",
    archivePrefix = "arXiv",
    reportNumber = "IHES-P-99-90, GRP-99-521",
    doi = "10.1103/PhysRevD.62.064015",
    journal = "Phys. Rev. D",
    volume = "62",
    pages = "064015",
    year = "2000"
}

@article{LIGOScientific:2025rsn,
    author = "Abac, A. G. and others",
    collaboration = "LIGO Scientific, VIRGO, KAGRA",
    title = "{GW231123: a Binary Black Hole Merger with Total Mass 190-265 $M_{\odot}$}",
    eprint = "2507.08219",
    archivePrefix = "arXiv",
    primaryClass = "astro-ph.HE",
    reportNumber = "DCC: P2500026-v6",
    month = "7",
    year = "2025"
}

@article{Chandra:2024dhf,
    author = "Chandra, Koustav",
    title = "{gwforge: a user-friendly package to generate gravitational-wave mock data}",
    eprint = "2407.21109",
    archivePrefix = "arXiv",
    primaryClass = "gr-qc",
    doi = "10.1088/1361-6382/ad9b68",
    journal = "Class. Quant. Grav.",
    volume = "42",
    number = "2",
    pages = "025003",
    year = "2025"
}

@article{Postnov:2014tza,
    author = "Postnov, Konstantin A. and Yungelson, Lev R.",
    title = "{The Evolution of Compact Binary Star Systems}",
    eprint = "1403.4754",
    archivePrefix = "arXiv",
    primaryClass = "astro-ph.HE",
    doi = "10.12942/lrr-2014-3",
    journal = "Living Rev. Rel.",
    volume = "17",
    pages = "3",
    year = "2014"
}

@article{LIGOScientific:2025snk,
    author = "Abac, A. G. and others",
    collaboration = "LIGO Scientific, VIRGO, KAGRA",
    title = "{Open Data from LIGO, Virgo, and KAGRA through the First Part of the Fourth Observing Run}",
    eprint = "2508.18079",
    archivePrefix = "arXiv",
    primaryClass = "gr-qc",
    reportNumber = "LIGO-P2500167",
    month = "8",
    year = "2025"
}

@article{Capote:2024rmo,
    author = "Capote, E. and others",
    title = "{Advanced LIGO detector performance in the fourth observing run}",
    eprint = "2411.14607",
    archivePrefix = "arXiv",
    primaryClass = "gr-qc",
    reportNumber = "LIGO-P2400256",
    doi = "10.1103/PhysRevD.111.062002",
    journal = "Phys. Rev. D",
    volume = "111",
    number = "6",
    pages = "062002",
    year = "2025"
}

@article{Cahillane:2022pqm,
    author = "Cahillane, Craig and Mansell, Georgia",
    title = "{Review of the Advanced LIGO Gravitational Wave Observatories Leading to Observing Run Four}",
    eprint = "2202.00847",
    archivePrefix = "arXiv",
    primaryClass = "gr-qc",
    reportNumber = "LIGO DCC P2100483",
    doi = "10.3390/galaxies10010036",
    journal = "Galaxies",
    volume = "10",
    number = "1",
    pages = "36",
    year = "2022"
}

@article{Harry:2017weg,
    author = "Harry, Ian and Calder{\'o}n Bustillo, Juan and Nitz, Alex",
    title = "{Searching for the full symphony of black hole binary mergers}",
    eprint = "1709.09181",
    archivePrefix = "arXiv",
    primaryClass = "gr-qc",
    reportNumber = "LIGO-DOCUMENT-P1700262, LIGO Document P1700262",
    doi = "10.1103/PhysRevD.97.023004",
    journal = "Phys. Rev. D",
    volume = "97",
    number = "2",
    pages = "023004",
    year = "2018"
}

@article{Pratten:2020ceb,
    author = "Pratten, Geraint and others",
    title = "{Computationally efficient models for the dominant and subdominant harmonic modes of precessing binary black holes}",
    eprint = "2004.06503",
    archivePrefix = "arXiv",
    primaryClass = "gr-qc",
    doi = "10.1103/PhysRevD.103.104056",
    journal = "Phys. Rev. D",
    volume = "103",
    number = "10",
    pages = "104056",
    year = "2021"
}

@article{LIGOScientific:2014pky,
    author = "Aasi, J. and others",
    collaboration = "LIGO Scientific",
    title = "{Advanced LIGO}",
    eprint = "1411.4547",
    archivePrefix = "arXiv",
    primaryClass = "gr-qc",
    doi = "10.1088/0264-9381/32/7/074001",
    journal = "Class. Quant. Grav.",
    volume = "32",
    pages = "074001",
    year = "2015"
}

@article{Damour:2010zb,
    author = "Damour, Thibault and Nagar, Alessandro and Trias, Miguel",
    title = "{Accuracy and effectualness of closed-form, frequency-domain waveforms for non-spinning black hole binaries}",
    eprint = "1009.5998",
    archivePrefix = "arXiv",
    primaryClass = "gr-qc",
    reportNumber = "LIGO-P1000099-V3",
    doi = "10.1103/PhysRevD.83.024006",
    journal = "Phys. Rev. D",
    volume = "83",
    pages = "024006",
    year = "2011"
}

@article{Damour:2007xr,
    author = "Damour, Thibault and Nagar, Alessandro",
    title = "{Faithful effective-one-body waveforms of small-mass-ratio coalescing black-hole binaries}",
    eprint = "0705.2519",
    archivePrefix = "arXiv",
    primaryClass = "gr-qc",
    doi = "10.1103/PhysRevD.76.064028",
    journal = "Phys. Rev. D",
    volume = "76",
    pages = "064028",
    year = "2007"
}

@article{Nagar:2020pcj,
    author = "Nagar, Alessandro and Riemenschneider, Gunnar and Pratten, Geraint and Rettegno, Piero and Messina, Francesco",
    title = "{Multipolar effective one body waveform model for spin-aligned black hole binaries}",
    eprint = "2001.09082",
    archivePrefix = "arXiv",
    primaryClass = "gr-qc",
    doi = "10.1103/PhysRevD.102.024077",
    journal = "Phys. Rev. D",
    volume = "102",
    number = "2",
    pages = "024077",
    year = "2020"
}

@article{Damour:2008gu,
    author = "Damour, Thibault and Iyer, Bala R. and Nagar, Alessandro",
    title = "{Improved resummation of post-Newtonian multipolar waveforms from circularized compact binaries}",
    eprint = "0811.2069",
    archivePrefix = "arXiv",
    primaryClass = "gr-qc",
    doi = "10.1103/PhysRevD.79.064004",
    journal = "Phys. Rev. D",
    volume = "79",
    pages = "064004",
    year = "2009"
}

@article{Pompili:2025cdc,
    author = "Pompili, Lorenzo and Maggio, Elisa and Silva, Hector O. and Buonanno, Alessandra",
    title = "{Parametrized spin-precessing inspiral-merger-ringdown waveform model for tests of general relativity}",
    eprint = "2504.10130",
    archivePrefix = "arXiv",
    primaryClass = "gr-qc",
    doi = "10.1103/ng8w-98sz",
    journal = "Phys. Rev. D",
    volume = "111",
    number = "12",
    pages = "124040",
    year = "2025"
}

@article{Ashton:2025xba,
    author = "Ashton, Gregory",
    title = "{Reconstructing and resampling: a guide to utilising posterior samples from gravitational wave observations}",
    eprint = "2510.11197",
    archivePrefix = "arXiv",
    primaryClass = "gr-qc",
    month = "10",
    year = "2025"
}

@article{Rover:2006ni,
    author = "Rover, Christian and Meyer, Renate and Christensen, Nelson",
    title = "{Bayesian inference on compact binary inspiral gravitational radiation signals in interferometric data}",
    eprint = "gr-qc/0602067",
    archivePrefix = "arXiv",
    doi = "10.1088/0264-9381/23/15/009",
    journal = "Class. Quant. Grav.",
    volume = "23",
    pages = "4895--4906",
    year = "2006"
}

@article{Cornish:2014kda,
    author = "Cornish, Neil J. and Littenberg, Tyson B.",
    title = "{BayesWave: Bayesian Inference for Gravitational Wave Bursts and Instrument Glitches}",
    eprint = "1410.3835",
    archivePrefix = "arXiv",
    primaryClass = "gr-qc",
    doi = "10.1088/0264-9381/32/13/135012",
    journal = "Class. Quant. Grav.",
    volume = "32",
    number = "13",
    pages = "135012",
    year = "2015"
}

@article{Dhani:2024jja,
    author = {Dhani, Arnab and V{\"o}lkel, Sebastian H. and Buonanno, Alessandra and Estelles, Hector and Gair, Jonathan and Pfeiffer, Harald P. and Pompili, Lorenzo and Toubiana, Alexandre},
    title = "{Systematic Biases in Estimating the Properties of Black Holes Due to Inaccurate Gravitational-Wave Models}",
    eprint = "2404.05811",
    archivePrefix = "arXiv",
    primaryClass = "gr-qc",
    doi = "10.1103/5pks-qz6b",
    journal = "Phys. Rev. X",
    volume = "15",
    number = "3",
    pages = "031036",
    year = "2025"
}

@unpublished{Gupta:2025xxx,
    author = "Gupta, Ish and others",
    title = "{Testing general relativity with amplitudes of subdominant gravitational-wave modes}",
    year = "2025",
    note = "{In preparation.}",
}

@article{Gerosa:2016sys,
    author = "Gerosa, Davide and Kesden, Michael",
    title = "{PRECESSION: Dynamics of spinning black-hole binaries with python}",
    eprint = "1605.01067",
    archivePrefix = "arXiv",
    primaryClass = "astro-ph.HE",
    doi = "10.1103/PhysRevD.93.124066",
    journal = "Phys. Rev. D",
    volume = "93",
    number = "12",
    pages = "124066",
    year = "2016"
}

@article{KAGRA:2021duu,
    author = "Abbott, R. and others",
    collaboration = "KAGRA, VIRGO, LIGO Scientific",
    title = "{Population of Merging Compact Binaries Inferred Using Gravitational Waves through GWTC-3}",
    eprint = "2111.03634",
    archivePrefix = "arXiv",
    primaryClass = "astro-ph.HE",
    reportNumber = "LIGO-P2100239 ; Data release: https://zenodo.org/record/5655785, LIGO-P2100239",
    doi = "10.1103/PhysRevX.13.011048",
    journal = "Phys. Rev. X",
    volume = "13",
    number = "1",
    pages = "011048",
    year = "2023"
}

@article{Gerosa:2017kvu,
    author = "Gerosa, Davide and Berti, Emanuele",
    title = "{Are merging black holes born from stellar collapse or previous mergers?}",
    eprint = "1703.06223",
    archivePrefix = "arXiv",
    primaryClass = "gr-qc",
    doi = "10.1103/PhysRevD.95.124046",
    journal = "Phys. Rev. D",
    volume = "95",
    number = "12",
    pages = "124046",
    year = "2017"
}

@article{Talbot:2017yur,
    author = "Talbot, Colm and Thrane, Eric",
    title = "{Determining the population properties of spinning black holes}",
    eprint = "1704.08370",
    archivePrefix = "arXiv",
    primaryClass = "astro-ph.HE",
    doi = "10.1103/PhysRevD.96.023012",
    journal = "Phys. Rev. D",
    volume = "96",
    number = "2",
    pages = "023012",
    year = "2017"
}

@article{Talbot:2018cva,
    author = "Talbot, Colm and Thrane, Eric",
    title = "{Measuring the binary black hole mass spectrum with an astrophysically motivated parameterization}",
    eprint = "1801.02699",
    archivePrefix = "arXiv",
    primaryClass = "astro-ph.HE",
    doi = "10.3847/1538-4357/aab34c",
    journal = "Astrophys. J.",
    volume = "856",
    number = "2",
    pages = "173",
    year = "2018"
}

@article{Chandra:2025ipu,
    author = "Chandra, Koustav and Calder{\'o}n Bustillo, Juan",
    title = "{Black-hole ringdown analysis with inspiral-merger informed templates and limitations of classical spectroscopy}",
    eprint = "2509.17315",
    archivePrefix = "arXiv",
    primaryClass = "gr-qc",
    reportNumber = "DCC-P2500566",
    month = "9",
    year = "2025"
}

@article{Isi:2021iql,
    author = "Isi, Maximiliano and Farr, Will M.",
    title = "{Analyzing black-hole ringdowns}",
    eprint = "2107.05609",
    archivePrefix = "arXiv",
    primaryClass = "gr-qc",
    reportNumber = "LIGO-P2100227",
    month = "7",
    year = "2021"
}

@article{Rashti:2024yoc,
    author = "Rashti, Alireza and Gamba, Rossella and Chandra, Koustav and Radice, David and Daszuta, Boris and Cook, William and Bernuzzi, Sebastiano",
    title = "{Binary black hole waveforms from high-resolution gr-athena++ simulations}",
    eprint = "2411.11989",
    archivePrefix = "arXiv",
    primaryClass = "gr-qc",
    doi = "10.1103/n5pz-qv3x",
    journal = "Phys. Rev. D",
    volume = "111",
    number = "10",
    pages = "104078",
    year = "2025"
}

@article{Saleem:2021nsb,
    author = "Saleem, Muhammed and Datta, Sayantani and Arun, K. G. and Sathyaprakash, B. S.",
    title = "{Parametrized tests of post-Newtonian theory using principal component analysis}",
    eprint = "2110.10147",
    archivePrefix = "arXiv",
    primaryClass = "gr-qc",
    doi = "10.1103/PhysRevD.105.084062",
    journal = "Phys. Rev. D",
    volume = "105",
    number = "8",
    pages = "084062",
    year = "2022"
}

@article{Agathos:2013upa,
    author = "Agathos, Michalis and Del Pozzo, Walter and Li, Tjonnie G. F. and Van Den Broeck, Chris and Veitch, John and Vitale, Salvatore",
    title = "{TIGER: A data analysis pipeline for testing the strong-field dynamics of general relativity with gravitational wave signals from coalescing compact binaries}",
    eprint = "1311.0420",
    archivePrefix = "arXiv",
    primaryClass = "gr-qc",
    doi = "10.1103/PhysRevD.89.082001",
    journal = "Phys. Rev. D",
    volume = "89",
    number = "8",
    pages = "082001",
    year = "2014"
}

@article{Isi:2019aib,
    author = "Isi, Maximiliano and Giesler, Matthew and Farr, Will M. and Scheel, Mark A. and Teukolsky, Saul A.",
    title = "{Testing the no-hair theorem with GW150914}",
    eprint = "1905.00869",
    archivePrefix = "arXiv",
    primaryClass = "gr-qc",
    reportNumber = "LIGO-P1900135",
    doi = "10.1103/PhysRevLett.123.111102",
    journal = "Phys. Rev. Lett.",
    volume = "123",
    number = "11",
    pages = "111102",
    year = "2019"
}

@article{Gamboa:2024hli,
    author = "Gamboa, Aldo and others",
    title = "{Accurate waveforms for eccentric, aligned-spin binary black holes: The multipolar effective-one-body model seobnrv5ehm}",
    eprint = "2412.12823",
    archivePrefix = "arXiv",
    primaryClass = "gr-qc",
    doi = "10.1103/jxrc-z298",
    journal = "Phys. Rev. D",
    volume = "112",
    number = "4",
    pages = "044038",
    year = "2025"
}

@article{Nagar:2024dzj,
    author = "Nagar, Alessandro and Gamba, Rossella and Rettegno, Piero and Fantini, Veronica and Bernuzzi, Sebastiano",
    title = "{Effective-one-body waveform model for noncircularized, planar, coalescing black hole binaries: The importance of radiation reaction}",
    eprint = "2404.05288",
    archivePrefix = "arXiv",
    primaryClass = "gr-qc",
    doi = "10.1103/PhysRevD.110.084001",
    journal = "Phys. Rev. D",
    volume = "110",
    number = "8",
    pages = "084001",
    year = "2024"
}

@article{Carullo:2019flw,
    author = "Carullo, Gregorio and Del Pozzo, Walter and Veitch, John",
    title = "{Observational Black Hole Spectroscopy: A time-domain multimode analysis of GW150914}",
    eprint = "1902.07527",
    archivePrefix = "arXiv",
    primaryClass = "gr-qc",
    doi = "10.1103/PhysRevD.99.123029",
    journal = "Phys. Rev. D",
    volume = "99",
    number = "12",
    pages = "123029",
    year = "2019",
    note = "[Erratum: Phys.Rev.D 100, 089903 (2019)]"
}

@article{Pompili:2023tna,
    author = "Pompili, Lorenzo and others",
    title = "{Laying the foundation of the effective-one-body waveform models SEOBNRv5: Improved accuracy and efficiency for spinning nonprecessing binary black holes}",
    eprint = "2303.18039",
    archivePrefix = "arXiv",
    primaryClass = "gr-qc",
    doi = "10.1103/PhysRevD.108.124035",
    journal = "Phys. Rev. D",
    volume = "108",
    number = "12",
    pages = "124035",
    year = "2023"
}

@article{Colleoni:2024knd,
    author = "Colleoni, Marta and Vidal, Felip A. Ramis and Garcia-Quiros, Cecilio and Akcay, Sarp and Bera, Sayantani",
    title = "{Fast frequency-domain gravitational waveforms for precessing binaries with a new twist}",
    eprint = "2412.16721",
    archivePrefix = "arXiv",
    primaryClass = "gr-qc",
    doi = "10.1103/PhysRevD.111.104019",
    journal = "Phys. Rev. D",
    volume = "111",
    number = "10",
    pages = "104019",
    year = "2025"
}

@article{Khalil:2023kep,
    author = "Khalil, Mohammed and Buonanno, Alessandra and Estelles, Hector and Mihaylov, Deyan P. and Ossokine, Serguei and Pompili, Lorenzo and Ramos-Buades, Antoni",
    title = "{Theoretical groundwork supporting the precessing-spin two-body dynamics of the effective-one-body waveform models SEOBNRv5}",
    eprint = "2303.18143",
    archivePrefix = "arXiv",
    primaryClass = "gr-qc",
    doi = "10.1103/PhysRevD.108.124036",
    journal = "Phys. Rev. D",
    volume = "108",
    number = "12",
    pages = "124036",
    year = "2023"
}

@article{vandeMeent:2023ols,
    author = "van de Meent, Maarten and Buonanno, Alessandra and Mihaylov, Deyan P. and Ossokine, Serguei and Pompili, Lorenzo and Warburton, Niels and Pound, Adam and Wardell, Barry and Durkan, Leanne and Miller, Jeremy",
    title = "{Enhancing the SEOBNRv5 effective-one-body waveform model with second-order gravitational self-force fluxes}",
    eprint = "2303.18026",
    archivePrefix = "arXiv",
    primaryClass = "gr-qc",
    doi = "10.1103/PhysRevD.108.124038",
    journal = "Phys. Rev. D",
    volume = "108",
    number = "12",
    pages = "124038",
    year = "2023"
}

@article{Ramos-Buades:2023ehm,
    author = "Ramos-Buades, Antoni and Buonanno, Alessandra and Estell{\'e}s, H{\'e}ctor and Khalil, Mohammed and Mihaylov, Deyan P. and Ossokine, Serguei and Pompili, Lorenzo and Shiferaw, Mahlet",
    title = "{Next generation of accurate and efficient multipolar precessing-spin effective-one-body waveforms for binary black holes}",
    eprint = "2303.18046",
    archivePrefix = "arXiv",
    primaryClass = "gr-qc",
    doi = "10.1103/PhysRevD.108.124037",
    journal = "Phys. Rev. D",
    volume = "108",
    number = "12",
    pages = "124037",
    year = "2023"
}

@article{Hawking:1971tu,
    author = "Hawking, S. W.",
    title = "{Gravitational radiation from colliding black holes}",
    doi = "10.1103/PhysRevLett.26.1344",
    journal = "Phys. Rev. Lett.",
    volume = "26",
    pages = "1344--1346",
    year = "1971"
}

@misc{LIGOScientific:2025tgr,
    author         = {Abac, A. G. and others},
    collaboration  = {LIGO Scientific, KAGRA, Virgo},
    eprint         = {25XX.XXXXX},
    note           = {(O4a TGR paper)},
}

@article{LIGOScientific:2021sio,
    author = "Abbott, R. and others",
    collaboration = "LIGO Scientific, VIRGO, KAGRA",
    title = "{Tests of General Relativity with GWTC-3}",
    eprint = "2112.06861",
    archivePrefix = "arXiv",
    primaryClass = "gr-qc",
    reportNumber = "LIGO-P2100275",
    doi = "10.1103/PhysRevD.112.084080",
    journal = "Phys. Rev. D",
    volume = "112",
    number = "8",
    pages = "084080",
    year = "2025"
}

@article{Mills:2020thr,
    author = "Mills, Cameron and Fairhurst, Stephen",
    title = "{Measuring gravitational-wave higher-order multipoles}",
    eprint = "2007.04313",
    archivePrefix = "arXiv",
    primaryClass = "gr-qc",
    doi = "10.1103/PhysRevD.103.024042",
    journal = "Phys. Rev. D",
    volume = "103",
    number = "2",
    pages = "024042",
    year = "2021"
}

@article{Gennari:2023gmx,
    author = "Gennari, Vasco and Carullo, Gregorio and Del Pozzo, Walter",
    title = "{Searching for ringdown higher modes with a numerical relativity-informed post-merger model}",
    eprint = "2312.12515",
    archivePrefix = "arXiv",
    primaryClass = "gr-qc",
    doi = "10.1140/epjc/s10052-024-12550-x",
    journal = "Eur. Phys. J. C",
    volume = "84",
    number = "3",
    pages = "233",
    year = "2024"
}

@article{Berti:2006wq,
    author = "Berti, Emanuele and Cardoso, Vitor",
    title = "{Quasinormal ringing of Kerr black holes. I. The Excitation factors}",
    eprint = "gr-qc/0605118",
    archivePrefix = "arXiv",
    doi = "10.1103/PhysRevD.74.104020",
    journal = "Phys. Rev. D",
    volume = "74",
    pages = "104020",
    year = "2006"
}

@article{LIGOScientific:2020tif,
    author = "Abbott, R. and others",
    collaboration = "LIGO Scientific, Virgo",
    title = "{Tests of general relativity with binary black holes from the second LIGO-Virgo gravitational-wave transient catalog}",
    eprint = "2010.14529",
    archivePrefix = "arXiv",
    primaryClass = "gr-qc",
    reportNumber = "LIGO-P2000091",
    doi = "10.1103/PhysRevD.103.122002",
    journal = "Phys. Rev. D",
    volume = "103",
    number = "12",
    pages = "122002",
    year = "2021"
}

@article{Ghosh:2021mrv,
    author = "Ghosh, Abhirup and Brito, Richard and Buonanno, Alessandra",
    title = "{Constraints on quasinormal-mode frequencies with LIGO-Virgo binary{\textendash}black-hole observations}",
    eprint = "2104.01906",
    archivePrefix = "arXiv",
    primaryClass = "gr-qc",
    reportNumber = "LIGO-P2100106",
    doi = "10.1103/PhysRevD.103.124041",
    journal = "Phys. Rev. D",
    volume = "103",
    number = "12",
    pages = "124041",
    year = "2021"
}

@unpublished{Grimaldi2025,
  author    = {Leonardo Grimaldi and others},
  title     = {in preparation},
  year      = {2025},
  note      = {Manuscript in preparation}
}

@article{Chiaramello:2025bhi,
    author = "Chiaramello, Danilo and Cibrario, Nicol{\`o} and Lange, Jacob and Chandra, Koustav and Gamba, Rossella and Bonino, Raffaella and Nagar, Alessandro",
    title = "{A parametrized model for gravitational waves from eccentric, precessing binary black holes: theory-agnostic tests of General Relativity with pTEOBResumS}",
    eprint = "2511.19593",
    archivePrefix = "arXiv",
    primaryClass = "gr-qc",
    month = "11",
    year = "2025"
}

@article{Andres-Carcasona:2025bni,
    author = "Andr{\'e}s-Carcasona, M. and Caneva Santoro, G.",
    title = "{No Love for black holes: tightest constraints on tidal Love numbers of black holes from GW250114}",
    eprint = "2512.01918",
    archivePrefix = "arXiv",
    primaryClass = "gr-qc",
    month = "12",
    year = "2025"
}

@article{Gamboa:2024imd,
    author = "Gamboa, Aldo and Khalil, Mohammed and Buonanno, Alessandra",
    title = "{Third post-Newtonian dynamics for eccentric orbits and aligned spins in the effective-one-body waveform model seobnrv5ehm}",
    eprint = "2412.12831",
    archivePrefix = "arXiv",
    primaryClass = "gr-qc",
    doi = "10.1103/rb1c-nx5f",
    journal = "Phys. Rev. D",
    volume = "112",
    number = "4",
    pages = "044037",
    year = "2025"
}

@article{Kumar:2025jio,
    author = "Kumar, Dhruv and Gupta, Ish and Sathyaprakash, Bangalore",
    title = "{Accelerating parameter estimation for parameterized tests of general relativity with gravitational-wave observations}",
    eprint = "2511.16879",
    archivePrefix = "arXiv",
    primaryClass = "gr-qc",
    month = "11",
    year = "2025"
}


\section*{Supplementary Material}

\paragraph*{\textbf{Analysis Setup:}} We analyse 8\,s of publicly available Advanced LIGO data around the event~\citep{LIGOScientific:2025snk},
sampled at 4096\,Hz, using \bilby\ with the \dynesty\ sampler and the \TEOBResumSDali\ waveform model
in four different configurations~\citep{Ashton:2018jfp, Speagle:2019ivv}: 
\begin{inparaenum}[(1)]
    \item non-eccentric, aligned-spin; 
    \item non-eccentric, precessing-spin; 
    \item eccentric, aligned-spin; and 
    \item eccentric, precessing-spin \ac{bbh} mergers.
\end{inparaenum} 
This model-selection step identifies the most probable configuration \textit{a priori}, thereby maximising the constraining power of subsequent deviation measurements~\footnote{By picking the most probable configuration, we shrink the allowed parameter space, which makes the subsequent measurements give tighter constraints.}.

Following~\citet{Romero-Shaw:2020owr}, we adopt isotropic priors for sky position and orientation, and a distance prior uniform in comoving volume and source time. Configuration-specific priors are used for intrinsic parameters--particularly spins, eccentricity, and mean anomaly--evaluated at a reference frequency of 13.3\,Hz.
To probe deviations from \ac{gr}, we use settings similar to those above, introducing uniform priors on the relevant deviation parameters as needed.

For our full-signal analyses, we use the Whittle likelihood, with the publicly available data power spectrum estimated on-source via a Bayesian approach that combines splines for broadband features and Lorentzians for narrow spectral lines~\citep{Cornish:2014kda}. For measuring the persistence of the subdominant \((4,\pm4)\) multipole in the post-inspiral signal, we instead employ a time-domain likelihood implemented in \bilby~\citep{Ashton:2018jfp};
further details are provided below. \\

\paragraph*{\textbf{The likelihood functions:}}
\label{sec:appendix-likelihood}

Bayesian inference libraries like \bilby\ model the evenly sampled time-domain detector data $\boldsymbol{d}$ as a linear superposition of a possible \ac{gw} signal $\boldsymbol{s}(\boldsymbol{\theta})$ and noise $\boldsymbol{n}$~\citep{Ashton:2018jfp}. Their goal is to identify a template $\boldsymbol{h}(\boldsymbol{\theta}\mid M)$ from a chosen waveform model $M$ (in our case, either \TEOBResumSDali\ or \bosch) that best represents the signal, such that the residual $\boldsymbol{r}=\boldsymbol{d}-\boldsymbol{h}(\boldsymbol{\theta}\mid M)$ follows the noise's statistical properties.

Generally, the noise $\boldsymbol{n}$ is assumed to be a zero-mean, wide-sense stationary Gaussian process with a covariance matrix \( C_{ik} = \langle n_i n_k \rangle \). Here, $n_i$ and $n_k$ are the noise samples at discrete times $t_i$ and $t_k$, respectively. If we further assume that the data are periodic with a time period $T$, then the covariance matrix in the Fourier domain is diagonal,
\begin{equation}
    \tilde{C}_{jj} = \langle |\tilde{n}_j|^2 \rangle = \frac{T}{2} S_j ,
\end{equation}
where $S_j$ is the one-sided noise power spectral density and
\begin{equation}
    \tilde{n}_j = \sum_{k=0}^{N-1} n_k \, e^{- 2 \pi i j k / N}
\end{equation}
is the discrete Fourier transform of $\boldsymbol{n}$ at frequency-bin index $j$, with $f_s = N/T$ the sampling frequency.

Under these assumptions, the single-detector log-likelihood ratio between the signal and noise assertion for the full analysis is~\citep{Finn:1992wt, Veitch:2009hd}
\begin{equation}\label{eq:likelihood}
    \ln \mathcal{L}^S_N = -\frac{1}{2} \langle \boldsymbol{h} \mid \boldsymbol{h} \rangle
    + \langle \boldsymbol{d} \mid \boldsymbol{h} \rangle ,
\end{equation}
where
\begin{equation}
    \langle \boldsymbol{a} \mid \boldsymbol{b} \rangle
    = \Re \sum_{j} \breve{a}_j^\ast \, \breve{b}_j
\end{equation}
is the noise-weighted inner product in the Fourier domain, and the whitened series is
\begin{equation}
    \breve{a}_j = \tilde{a}_j \sqrt{\frac{4}{T S_j}} .
\end{equation}

Equivalently, we may express the likelihood in the discrete time domain using the noise covariance matrix $C_{ik}$~\citep{Isi:2021iql}.  
In our analysis, we compute $C$ from the autocorrelation function obtained by inverse Fourier transforming the PSD $S_j$.
Assuming wide-sense stationarity, $C$ is Toeplitz and positive definite; we perform a Cholesky decomposition,
\begin{equation}
    C = L \, L^{\mathrm{T}} ,
\end{equation}
to define whitened time-domain quantities
\begin{equation}
    \bar{a}_i = \sum_{k} (L^{-1})_{ik} a_k ,
\end{equation}.

Restricting the analysis to a time window $[t_\mathrm{start},\,t_\mathrm{end}]$, we define the \emph{time-limited} noise-weighted inner product
\begin{equation}
    \langle \boldsymbol{a} \mid \boldsymbol{b} \rangle_{[t_\mathrm{start},\,t_\mathrm{end}]}
        = \sum_{i} \bar{a}_i \, \bar{b}_i ,
\end{equation}
which we use to evaluate the log-likelihood ratio defined in Eq.~\eqref{eq:likelihood} to probe the temporal structure of
\((4,\pm4)\) modes~\citep{Chandra:2025ipu}. 

Instead of performing a new stochastic sampling for this restricted window, we recompute the likelihood using Eq.~\eqref{eq:likelihood} by \emph{resampling} from the existing \ac{imr} posteriors~\citep{Rover:2006ni, Ashton:2025xba}. 
This means the samples from the \ac{imr} analysis act as draws from a \emph{proposal distribution}, while the restricted time-domain likelihood defines a new \emph{target distribution}. 
The ratio of these two, expressed as a weight factor for each sample, reweights both the \ac{imr} posterior and the evidence, providing an approximation to the restricted-domain posterior without rerunning the full sampler.

This method is computationally efficient but comes with an important caveat: the true posterior from a restricted time-domain analysis is generally broader than that from the full \ac{imr} run, since the restricted likelihood omits part of the signal and weakens correlations among parameters.
Its validity, therefore, depends on the proposal distribution adequately covering the high-probability region of the target distribution; any coverage gaps or sampling bias in the \ac{imr} analysis will be inherited by the resampled posterior. Given that our primary interest is the relative evidence, this procedure offers a fast, low-cost approximation. To ensure its robustness, we independently cross-checked the results by carrying out full analyses for $\Delta t^\mathrm{geo}_\mathrm{start}/M \in \{-5,\,0,\,5,\,10\}$, confirming consistency with the resampled estimates.  \\

\begin{figure*}
    \centering
    \includegraphics[width=0.49\textwidth]{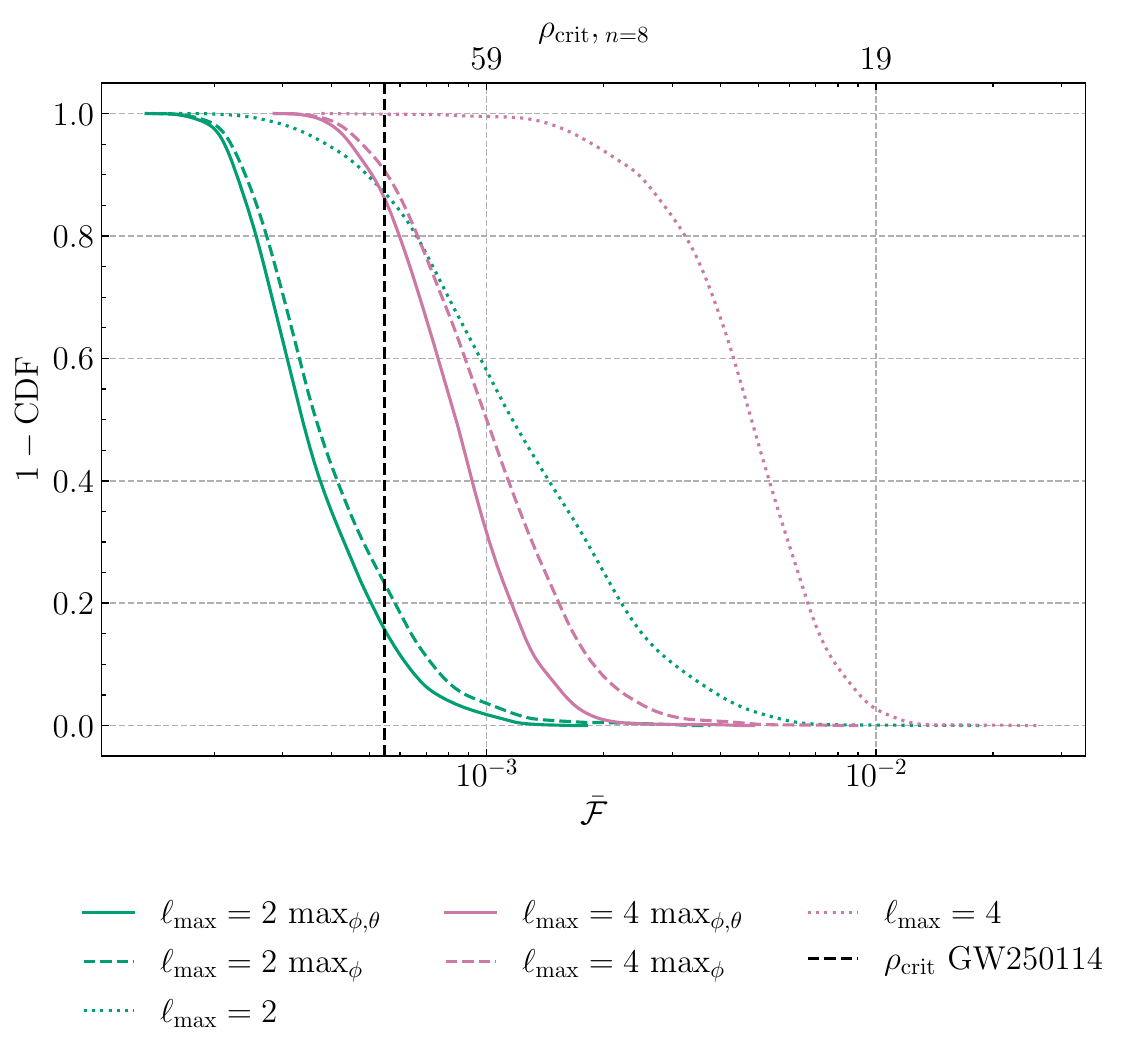} 
    \includegraphics[width=0.49\textwidth]{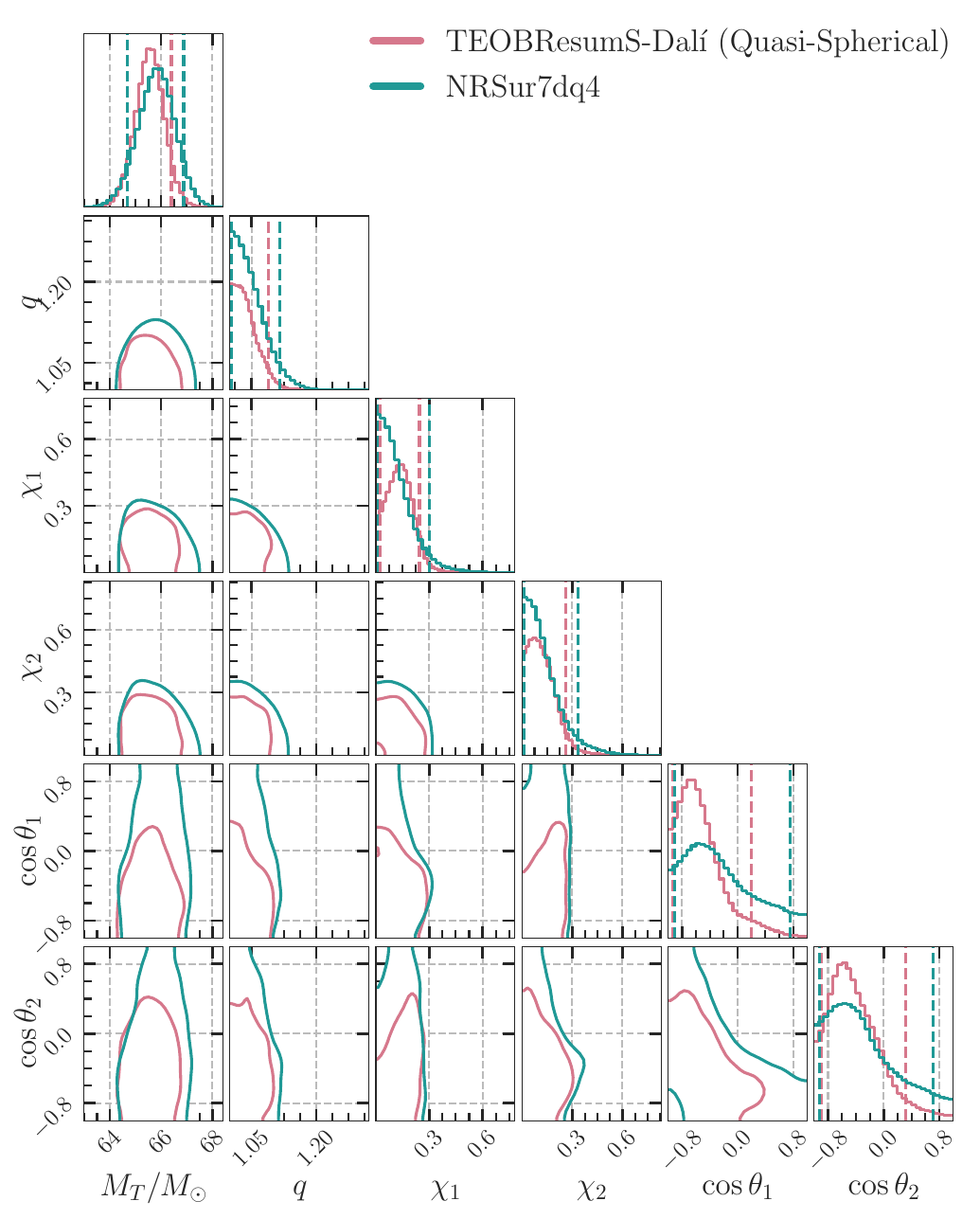}
    \caption{
        \textbf{Left panel:} Unfaithfulness \(\bar{\mathcal{F}}\) between the \texttt{NRSur7dq4} and \TEOBResumSDali\ waveform model
        for binary parameters consistent with the posteriors. The dashed and dotted lines show the mismatches 
        obtained when no in-plane rotation ($\theta$) and azimuth ($\phi$) minimisations are performed. Different colors correspond to
        mismatches obtained considering models up to $\ell = 2$ (green) and $\ell=4$ (pink).
        The vertical black line marks the critical threshold below which the models are indistinguishable,
        ensuring unbiased inference at the signal's \ac{snr}.
        \textbf{Right panel:} Joint Posterior of mass and spin parameters for GW250114 inferred using the \TEOBResumSDali\ (red) and \texttt{NRSur7dq4} (teal) waveform model. The posteriors broadly agree. 
    }
    \label{fig:mismatch}
\end{figure*}

\paragraph*{\textbf{Comparison against surrogate models:}}
\label{sec:mismatches}

To assess the robustness of our \ac{gr} baseline model, we compute mismatches between \TEOBResumSDali\ and \texttt{NRSur7dq4}~\citep{Varma:2019csw}, 
a quasi-spherical multipolar model built by interpolating \ac{nr} simulations and widely regarded as the most \ac{nr}-faithful waveform currently available.
We use the sky-maximized detection statistics of~\citet{Harry:2017weg}, maximized over time and reference azimuth, as well as
over an additional in-plane rotation angle~\cite{Pratten:2020ceb}. We assume a lower frequency cutoff of \(20\) Hz and an upper cutoff of \(900\) Hz, 
employing the Advanced LIGO design sensitivity curve~\citep{LIGOScientific:2014pky}.
We find mismatches of \(\bar{\mathcal{F}} \sim 10^{-4}-10^{-3}\) (\(10^{-3}-10^{-2}\)) between the two models
when considering only the $\ell =2$ (including up to $\ell=4$) modes for systems drawn from the posterior of \event~(see left panel of Fig.~\ref{fig:mismatch}).
These numbers translate into critical \ac{snr} thresholds of \(\rho_\mathrm{crit} \sim 70-100\) (\(20-30\)) below which the two models
should be indistinguishable~\cite{Damour:2010zb}. Therefore, while the $\ell=2$ modes of the two models are mostly indistinguishable for \event, the inclusion of higher-order modes
may in principle lead to measurable differences given the signal's high \ac{snr}. 

As a further validation, we repeated the \ac{pe} using \texttt{NRSur7dq4}, and compared the results with those obtained from \TEOBResumSDali.
The right panel of Figure~\ref{fig:mismatch} shows the posterior distributions of the mass and spin parameters, which are in broad agreement across the two
models.\\

\paragraph*{\textbf{Toy Monte Carlo Proof of GW250114's Isolated-Binary Merger Origin}:}
\label{sec:appendix-pop}
To demonstrate that GW250114 is less likely to have formed through a hierarchical merger process, we conduct a simple Monte Carlo study informed by the GWTC–3 population estimates~\citep{KAGRA:2021duu}. We generate an initial \ac{bbh} population by drawing component masses $m_k$ from the \texttt{Power Law + Peak} model, and spin magnitudes $\chi_k$ and tilt angles $\theta_k$ from the \texttt{Default} spin model~\citep{Talbot:2017yur, Talbot:2018cva}. We label the \acp{bh} in this sample as ``1G'', while noting that the observed GWTC–3 population may itself contain hierarchically assembled remnants; thus, this population does not represent a \textit{pure} 1G population aka \acp{bh} formed directly via stellar-collapse.

We merge a fraction of this 1G+1G population to form second-generation (2G) \acp{bh}. Their final masses and spins are computed using NR-informed fits \citep{Hofmann:2016yih, Jimenez-Forteza:2016oae}, and their recoil speeds are obtained using fits from \citet{Gerosa:2016sys}. For subsequent steps, we retain only those remnants whose recoil speeds are below $500\,\mathrm{km\,s^{-1}}$, a threshold achievable only in a few host environments (e.g., AGN discs~\citep{Gerosa:2017kvu}); thus, our results provide optimistic upper bounds for hierarchical growth.

Next, we construct mixed-generation 1G+2G and 2G+2G binaries by pairing black holes from the 1G and 2G samples with equal probability. We draw their spin orientations from an isotropic distribution, as expected because dynamical pairing leads to random spin alignment. We subsequently merge these binaries to produce third-generation (3G) \acp{bh} and explore 1G+3G, 2G+3G, and 3G+3G pairings in a similar fashion.

For each population (1G+1G, 1G+2G, 2G+2G, ...), we compute the probability using GW250114's posteriors, obtained under the quasi-spherical \acp{bbh} interpretation, to belong to a given population. We find
\[
P_{\mathrm{1G+1G}} \sim 0.967, \quad
P_{\mathrm{1G+2G}} \sim 0.032, \quad
P_{\mathrm{2G+2G}} \sim 0.001.
\]
The much lower probabilities for hierarchical pairings arise because near-equal-mass, low-spin binaries ($q\to1$, $\chi_{1,2}\lesssim0.2$) tend to produce remnants with $\chi_f\gtrsim0.5$, largely independent of spin orientation. Probabilities fall further for 1G+3G, 2G+3G, and 3G+3G mergers. We repeat this analysis multiple times and obtain similar numbers every time.

In summary, this simple experiment strongly favours a 1G+1G origin for GW250114. While idealised—neglecting detection probabilities, realistic retention fractions, pairing biases, and environment-specific dynamics—our assumptions already maximise the probability of hierarchical assembly. \\

\end{document}